\documentclass[reqno]{amsart}

\usepackage{amssymb,epsfig,graphics}
\usepackage{amsmath}
\usepackage{amscd}
\usepackage{verbatim}
\input xypic

\theoremstyle{definition}

\newtheorem*{example}{Example}

\def\C{\mathbb C}
\def\R{\mathbb R}

\def\c{\cdot}

\def\o{\omega}

\begin{document}

\title[Branching of $B_n$, $C_n$, and $D_n$ orbits]
{Branching rules for Weyl group orbits\\ of simple Lie algebras $B_n$, $C_n$ and $D_n$}

\author{M. Larouche and J. Patera}
\address{Centre de recherches math\'ematiques,
         Universit\'e de Montr\'eal,
         C.P.~6128 Centre-ville,
         Montr\'eal, H3C\,3J7, Qu\'ebec, Canada}
\email{larouche@dms.umontreal.ca, patera@crm.umontreal.ca}

\date{\today}
 \begin{abstract}
The orbits of Weyl groups $W(B_n)$, $W(C_n)$ and $W(D_n)$ of the simple Lie algebras $B_n$, $C_n$ and $D_n$ are reduced to the union of the orbits of Weyl groups of the maximal reductive subalgebras of $B_n$, $C_n$ and $D_n$. Matrices transforming points of $W(B_n)$, $W(C_n)$ and $W(D_n)$ orbits into points of subalgebra orbits are listed for all cases $n\leq8$ and for the infinite series of algebra-subalgebra pairs.
$B_n\supset B_{n-1}\times U_1$, \  
$B_n\supset D_n$, \  
$B_n\supset B_{n-k}\times D_k$, \  
$B_n\supset A_1$, \  
$C_n\supset C_{n-k}\times C_k$, \  
$C_n\supset A_{n-1}\times U_1$, \  
$D_n\supset A_{n-1}\times U_1$, \  
$D_n\supset D_{n-1}\times U_1$, \  
$D_n\supset B_{n-1}$, \  
$D_n\supset B_{n-k-1}\times B_k$, \
$D_n\supset D_{n-k}\times D_k$. \  
Numerous special cases and examples are shown.
\end{abstract}
\maketitle

\section{Introduction}
This paper is a continuation of \cite{LNP}, in which the analogous problem for Lie algebras $A_n$ of the special linear group $SL(n+1,\C)$ was considered. Here the problem is considered for simple Lie algebras $B_n$ and $D_n$ of orthogonal groups $O(2n+1)$ and $O(2n)$ respectively, and for the simple Lie algebra $C_n$ of the symplectic group $Sp(2n)$. 

The motivation for the present paper is the same as in \cite{LNP}. There are four important points to note: firstly, orbit branching rules are implicitly required for the computation of branching rules of representations of the same Lie algebra-subalgebra pairs. 
Hence, projection matrices, an essential part of the method in \cite{LNP}, are used as the main tool in the paper. Secondly, it turns out that, for any extensive computation with finite-dimensional representations of simple Lie algebras such as branching rules, the decomposition of tensor products of representations, or discrete Fourier analysis, it is impracticable to avoid decomposing the problem into several subproblems for orbits involved. This is because the dimensions of representations increase without bound, while Weyl group orbits are of finite size in all cases, their size always being a divisor of the order of the corresponding Weyl group. 

An important property as yet unexploited in applications is the fact that Weyl group orbit points do not need to belong to a lattice. Weyl group orbits that are not on the corresponding weight lattice retain most of the valuable properties of orbits that are on the lattice. In particular, branching rules remain valid even if the coordinates of the orbit points are irrational numbers. Recent interest in special functions defined by Weyl group orbits \cite{KP1,KP2} is based on  knowledge of orbit properties.

It should also be noted that Lie algebras of type $B_n$, $C_n$ and $D_n$ are amenable for a different choice of basis than that used in this paper, namely the orthonormal basis. For some problems, this choice may offer a simplifying advantage in terms of computation. We refrain from using it here in favour of the non-orthogonal root and weight bases, because these offer a remarkable uniformity of computation methods for semisimple Lie algebras of all types.

The paper contains projection matrices for all cases of maximal inclusion for Lie algebras of types $B_n$, $C_n$, and $D_n$ for ranks $n\leq8$, with examples of branching rules for specific orbits. In addition, projection matrices and examples of branching rules for infinite series of selected cases are given. Included are all cases where a maximal reductive subalgebra is of the same rank as $B_n$, $C_n$, and $D_n$.

Branching rules for Weyl group orbits of exceptional simple Lie algebras $E_6$, $E_7$, $E_8$, $F_4$, and $G_2$ are found in \cite{MPR} among many other results.

The branching rules for $W(L)\supset W(L')$, where $L'$ is a maximal reductive subalgebra of $L$, is a linear transformation between Euclidean spaces $\R^n\rightarrow\R^{n'},$ where $n$ and $n'$ are the ranks of $L$ and $L'$ respectively. The branching rules are unique, unlike transformations of individual orbit points, which depend on the relative choice of bases. We provide the linear transformation in the form of an $n'\times n$ matrix, the `projection matrix'. A suitable choice of bases allows one to obtain integer matrix elements in all the projection matrices listed here. Note that we use Dynkin notations and numberings for roots, weights and diagrams.

The method we use here is an extension of the method used in \cite{MPR,McP,MPS,PSan} for the computation of reductions of representations of simple Lie algebras to representations of their maximal semisimple subalgebras. Orbit-orbit branching rules have been discussed for one of the first times in the literature in \cite{MPR}. They were then addressed in \cite{GPS, ST2, ST1}, where specific methods were developed for different algebra-subalgebra pairs. The main advantage of the projection matrix method is its uniformity, as it can be used for any algebra-subalgebra pair. We include here, as we did in \cite{LNP}, all the cases when the maximal reductive subalgebra is non-semisimple, i.e when it contains the 1-parametric ideal denoted here $U_1$. 

It should be underlined that each of the numerous examples of orbit branching rules shown here is valid for an infinity of cases. For example, an orbit labeled by $(a,0,\dots,0)$, refers to an uncountable number of orbits with $0<a\in\R$. Orbits that do not belong to a weight lattice should be of importance in Fourier analysis when considering Fourier integrals rather than Fourier series. 

The number attached to each representation of a simple Lie algebra and called the second degree index is an invariant of the representation which has been occasionally used in applications \cite{Slansky}. Its useful properties remain valid also for Weyl group orbits. The index of a semisimple subalgebra in a simple Lie algebra is an invariant of all branching rules for a fixed algebra-subalgebra pair. It was introduced in \cite{Dynkin}, see Equation (2.26). It is defined using the second degree indices of representations. We give its value for all our cases, but its properties would merit further investigation, particularly when the orbit points are off the weight lattices.

\section{Preliminaries}

Finite groups generated by reflections in an $n$-dimensional real Euclidean space~$\R^n$ are commonly known as finite Coxeter groups \cite{H}. Finite Coxeter groups are split into two classes: crystallographic and non crystallographic groups. Crystallographic groups are often referred to as Weyl groups of semisimple Lie groups or Lie algebras. In $\R^n$ they are the symmetry groups of root lattices of the simple Lie groups. There are four infinite series (as to the admissible values of rank $n$) of such groups, namely $A_n$, $B_n$, $C_n$, $D_n$, and five isolated exceptional groups of ranks 2, 4, 6, 7, and 8. The non crystallographic finite Coxeter groups are the symmetry groups of regular $2D$ polygons (the dihedral groups), with two exceptional groups, one of rank 3 -- the icosahedral group of order 120 -- and one of rank 4, which is of order $120^2$.

We consider orbits of the Weyl groups $W(B_n)$, $W(C_n)$ and $W(D_n)$ of the simple Lie algebras of type $B_n$, $n\geq2$,  $C_n$, $n\geq2$ and $D_n$, $n\geq 4$, respectively (Fig.~1). The order of such Weyl groups is $2^nn!$ for $W(B_n)$ and $W(C_n)$, while it is $2^{n-1}n!$ for $W(D_n)$. An orbit $W_\lambda$ of the Weyl group $W(L)$, where $L$ is of rank $n$, is a finite set of distinct points in $\R^n$, all equidistant from the origin, obtained from a single point $\lambda\in\R^n$ by application of $W$ to $\lambda$. Hence, an orbit of $W(B_n)$ or $W(C_n)$ contains at most $2^nn!$ points, and an orbit $W_\lambda$ of $W(D_n)$ contains at most $2^{n-1}n!$ points. 

Consider the pair $W(L)\supset W(L')$, where $L'$ is a maximal reductive subalgebra of a simple Lie algebra $L$. The orbit reduction is a linear transformation $\R^n\rightarrow\R^{n'}$, where $n'$ is the rank of $L'$. Hence the orbit reduction problem is solved when the $n' \times n$ matrix $P$ is found with the property that points of any orbit of $W(L)$ are projected by $P$ into points of the corresponding orbits of $W(L')$. Computation of the branching rule for a specific orbit of $W(L)$ amounts to applying $P$ to the points of the orbit, and to sorting out the projected points into a sum (union) of orbits of $W(L')$. 

Typically the result of the reduction of an orbit $W_\lambda$ of $W(L)$ is a union of several orbits of $W(L')$. Geometrically the points of $W_\lambda$ can be understood as vertices of a polytope in $\R^n$. A union of several obits is then an onion-like formation of concentric polytopes \cite{HLP}.

The projection matrix $P$ is calculated from one known branching rule. The classification of maximal reductive subalgebras of simple Lie algebras \cite{Dynkin, BdeS} provides the information to find that branching rule. The projection matrix is then obtained using the weight systems of the representations, by requiring that weights of $L$ be transformed by $P$ to weights of $L'$. Since any ordering of the weights is admissible, the projection matrix is not unique. We choose the natural lexicographical ordering of the weights. The projection matrix obtained can then be used to project points of any orbit of $W(L)$ into points of orbits of $W(L')$. At the end of this section, we consider an example of the construction of a projection matrix for the case $W(B_3)\supset W(G_2)$.

To compute the branching rule for a specific orbit of $W(L)$, all the points of that orbit are listed and then multiplied by the projection matrix. A standard method to calculate points of an orbit of any finite Coxeter group is given in \cite{HLP}, where the points are given in the corresponding basis of fundamental weights, called the $\o$-basis. All of the orbits appearing here are given in the $\o$-basis of the corresponding group, linked to the basis of simple roots by the Cartan matrix of the group. Since every orbit contains precisely one point with nonnegative coordinates in the $\o$-basis, the orbit can be identified by that point, called the dominant point of the orbit. Hence when referring to an orbit, one does not have to list all of the points it contains. The example at the end of this section illustrates the actual computation of branching rules for the case $W(B_3)\supset W(G_2)$.

The Weyl group of the one-parameter Lie algebra $U_1$ is trivial, consisting of the identity element only. Its irreducible representations are all 1-dimensional, hence its orbits consist of one element. They are labeled by integers, which can also take negative values. The symbol $(k)$ may stand for either the orbit $\{k,-k\}$ of $W(A_1)$, or for the $W(U_1)$ orbit of one point $\{k\}$. Distinction should be made from the context. Since we are working with orbits of the Weyl group of $U_1$ and the compactness of the Lie group is of no interest to us here, we can allow the orbits of $W(U_1)$ to take real values.

The second degree index for weight systems of irreducible finite dimensional representations of compact semisimple Lie groups was defined in \cite{PSW}. It was then introduced for individual orbits in \cite{HLP}. The second degree index $I_\lambda^{(2)}$ of the orbit $W_\lambda$ is 
\begin{equation}
 I_\lambda^{(2)}=\sum_{\mu\in W_\lambda}(\mu|\mu)=(\lambda|\lambda)|W_\lambda|\,,\notag
\end{equation}
where $|W_\lambda|$ is the size of the orbit and $(\c|\c)$ is the standard inner product of $\R^n$. The second equality comes from the fact that all points of $W_\lambda$ are equidistant from the origin. If $W_{\lambda_1}$ and $W_{\lambda_2}$ are two orbits of $W$, then the index of their sum (or union) and the index of their product are given by  
\begin{align}
I_{\lambda_1 + \lambda_2}^{(2)}
     &=I_{\lambda_1}^{(2)} + I_{\lambda_2}^{(2)} \notag \\
I_{\lambda_1\times \lambda_2}^{(2)}
     &=I_{\lambda_1}^{(2)}\,|W_{\lambda_2}|
     +I_{\lambda_2}^{(2)}\,|W_{\lambda_1}|\\
     &=|W_{\lambda_1}|\,|W_{\lambda_2}|\left(
     (\lambda_1|\lambda_1)+(\lambda_2|\lambda_2)\right)\,.
\end{align}

Simple calculations show that if $W^1_{\lambda_1}$ and $W^2_{\lambda_2}$ are two orbits of two different Weyl groups $W^1$ and $W^2$, the second degree index of the orbit $\lambda_1\times \lambda_2$ of $W^1\times W^2$ is also given by (1) and (2).

For a fixed pair $W(L)\supset W(L')$ of Weyl groups of an algebra $L$ and its semisimple subalgebra $L'$, the ratio of second degree indices is invariant and is called the index of $L'$ in $L$. For any orbit $W(L)_\lambda$ reduced to the sum of orbits  $\displaystyle\sum\limits_{\mu} W(L')_{\mu}$, there exists a positive number $\gamma=\gamma_{L,L'}$ such that
$$
I_{\lambda}^{(2)} = \gamma_{L,L'}\displaystyle\sum\limits_{\mu} I_{\mu}^{(2)} \,.
$$ 
We give that number $\gamma_{L,L'}$ for all such pairs of Weyl groups $W(L)\supset W(L')$.

To alleviate notation, we will simply write $L$ instead of $W (L)$ to refer to the Weyl group of the Lie algebra $L$, and $\lambda$ instead of $W_\lambda$ to refer to the orbit of the dominant point $\lambda$ of the Weyl group $W$. Subsequently dots in a matrix denote zero matrix elements.

Let us finally consider an example to illustrate how to construct a projection matrix and how to calculate a particular branching rule.
\begin{example}\

Consider the case of $B_3\supset G_2$ of subsection 3.2. From the classification of maximal reductive subalgebras, we know that the lowest orbit of $B_3$, the orbit of the dominant point $(1,0,0)$, contains 6 points  and is projected onto the $G_2$-orbit of the point $(0,1)$, that also contains 6 points. We order the points of the two orbits, and require that points of the first one be transformed into points of the second one in the following manner :
\begin{alignat*}{3}
&(1,0,0)\mapsto (0,1),\quad &&(\!{-}1,1,0)\mapsto (1,\!{-}1),\qquad &(0,\!{-}1,2)\mapsto (\!{-}1,2),\\
&(0,1,\!{-}2)\mapsto (1,\!{-}2),\quad &&(1,\!{-}1,0)\mapsto (\!{-}1,1),\quad &(\!{-}1,0,0)\mapsto (0,\!{-}1).
\end{alignat*}

Writing the points as column matrices, the projection matrix of subsection 3.2 is obtained from the first three. Proceeding one column at a time, we have
\begin{alignat*}{3}
\left(\begin{smallmatrix} 0 & * & * \\ 1 & * & * \end{smallmatrix}\right)
           \left(\begin{smallmatrix} 1 \\ 0 \\ 0 \end{smallmatrix}\right)
           &=\left(\begin{smallmatrix} 0 \\ 1  \end{smallmatrix}\right),&\qquad
\left(\begin{smallmatrix} 0 & 1 & * \\ 1 & 0 & * \end{smallmatrix}\right)
           \left(\begin{smallmatrix}\!\!{-}1 \\ 1 \\ 0 \end{smallmatrix}\right)
           &=\left(\begin{smallmatrix}1 \\ \!\!{-}1  \end{smallmatrix}\right),&\qquad
\left(\begin{smallmatrix} 0 & 1 & 0 \\ 1 &0 & 1\end{smallmatrix}\right)
           \left(\begin{smallmatrix} 0 \\ \!\!{-}1 \\ 2 \end{smallmatrix}\right)
           &=\left(\begin{smallmatrix} \!\!{-}1 \\ 2  \end{smallmatrix}\right),
\end{alignat*}
where stars denote the entries that are still to be determined. The matrix ${\mbox {\it P}{=}\left(\!\begin{smallmatrix} 0 & 1 & 0 \\ 1 &0 & 1\end{smallmatrix}\!\right)}$ then automatically transforms the three last points of the $B_3$-orbit as required. This matrix can then be used for projecting points of any $B_3$-orbit. For example, to calculate the reduction of the $B_3$-orbit of $(0,2,0)$, one has to write the coordinates of the 12 points of the orbit as column vectors :
\begin{gather}\label{B3orbit}
\begin{gathered}
\left(\begin{smallmatrix} 0 \\ 2 \\ 0  \end{smallmatrix}\right),
\left(\begin{smallmatrix} 2 \\ \!{-}2 \\ 4  \end{smallmatrix}\right),
\left(\begin{smallmatrix} \!{-}2 \\ 0 \\ 4  \end{smallmatrix}\right),
\left(\begin{smallmatrix} 2 \\ 2 \\ \!{-}4  \end{smallmatrix}\right),
\left(\begin{smallmatrix} \!{-}2 \\ 4 \\\!{-}4  \end{smallmatrix}\right),
\left(\begin{smallmatrix} 4 \\\!{-}2 \\ 0  \end{smallmatrix}\right),
\\
\left(\begin{smallmatrix} \!{-}4 \\2 \\ 0  \end{smallmatrix}\right),
\left(\begin{smallmatrix}2 \\ \!{-}4 \\4  \end{smallmatrix}\right),
\left(\begin{smallmatrix}\!{-}2 \\ \!{-}2 \\ 4  \end{smallmatrix}\right),
\left(\begin{smallmatrix} 2 \\ 0 \\ \!{-}4  \end{smallmatrix}\right),
\left(\begin{smallmatrix} \!{-}2 \\ 2 \\ \!{-}4  \end{smallmatrix}\right),
\left(\begin{smallmatrix}0 \\ \!{-}2 \\ 0  \end{smallmatrix}\right).
\end{gathered}
\end{gather}
Multiplying each of the points of \eqref{B3orbit} by the matrix $P$, one gets the points of the $G_2$-orbits written as column vectors. Rewriting them in the horizontal form, we have the set of projected points. To distribute the points into individual orbits, one only has to select the dominant points (no negative coordinates) because they represent the orbits that are present. Hence one gets the following branching rule for that case:
$$
(0,2,0) \supset (2,0)+(0,2)\,.
$$

\end{example}

\begin{figure}
 \hspace{-5cm}
\parbox{.6\linewidth}{\setlength{\unitlength}{2pt}
\def\kr{\circle{4}}
\def\cr{\circle*{4}}
\thicklines
\begin{picture}(180,75)

\put(30,69){\makebox(0,0){${B_n,\quad n\geq 2}$}}
\put(20,50){\kr}\put(19,44){\tiny1}
\put(30,50){\kr}\put(29,44){\tiny2}\put(30,52){\line(0,1){4}}
\put(30,58){\kr}\put(29.3,56.7){$\cdot$}\put(24.5,57){\tiny0}
\put(40,50){\kr}\put(39,44){\tiny3}
\put(47,50){$\dots$}
\put(60,50){\kr}\put(56,44){\tiny{$n-1$}}
\put(70,50){\cr}\put(69,44){\tiny{$n$}}
\put(22,50){\line(1,0){6}}
\put(32,50){\line(1,0){6}}
\put(42,50){\line(1,0){4}}
\put(54,50){\line(1,0){4}}
\put(61.6,51){\line(1,0){7}}
\put(61.6,49){\line(1,0){7}}

\put(105,69){\makebox(0,0){${C_n,\quad n\geq 2}$}}
\put(95,50){\kr}\put(94.3,48.7){$\cdot$}\put(94,44){\tiny0}
\put(105,50){\cr}\put(104,44){\tiny1}
\put(115,50){\cr}\put(114,44){\tiny2}
\put(122,50){$\dots$}
\put(135,50){\cr}\put(131,44){\tiny{$n-1$}}
\put(145,50){\kr}\put(144,44){\tiny{$n$}}
\put(107,50){\line(1,0){6}}
\put(116.8,50){\line(1,0){4}}
\put(129,50){\line(1,0){4}}
\put(96.6,51){\line(1,0){7}}
\put(96.6,49){\line(1,0){7}}
\put(136.4,51){\line(1,0){7}}
\put(136.4,49){\line(1,0){7}}

\put(30,34){\makebox(0,0){${D_n,\quad n\geq 4}$}}
\put(20,15){\kr}\put(19,9){\tiny1}
\put(30,15){\kr}\put(29,9){\tiny2}
\put(30,23){\kr}\put(29.3,21.7){$\cdot$}\put(24.5,22){\tiny0}
\put(40,15){\kr}\put(39,9){\tiny3}
\put(60,15){\kr}\put(54,9){\tiny{$n-3$}}
\put(70,15){\kr}\put(66,9){\tiny{$n-2$}}
\put(80,15){\kr}\put(78,9){\tiny{$n-1$}}
\put(70,23){\kr}\put(73.5,22){\tiny{$n$}}
\put(22,15){\line(1,0){6}}
\put(32,15){\line(1,0){6}}
\put(42,15){\line(1,0){4}}
\put(47,15){$\dots$}
\put(54,15){\line(1,0){4}}
\put(62,15){\line(1,0){6}}
\put(72,15){\line(1,0){6}}
\put(30,17){\line(0,1){4}}
\put(70,17){\line(0,1){4}}

\end{picture}
}
\caption{The Coxeter-Dynkin diagrams of types $B_n$, $C_n$ and $D_n$  are shown. The circular nodes stand for the simple roots, with the convention that open (resp. filled) circles indicate long (resp. short) roots. The dotted node is the negative highest root denoted $\alpha_0$. A link between a pair of roots indicates that the roots are not orthogonal. The Dynkin numbering of the nodes is shown.}
\label{diagrams}
\end{figure}
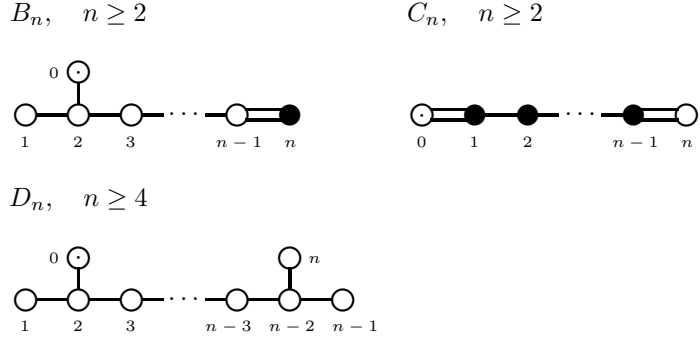

\section{Reduction of orbits of the Weyl group of $B_n$}
In this section we first consider all cases of dimension (rank of the Lie algebra) up to 8. In the last subsection, 3.8, we present infinite series of cases which occur for all values of rank starting from a lowest one. For each case, the projection matrix is given, together with examples of the corresponding reductions/branching rules. For cases involving Weyl groups of a simple algebra $L$ and a maximal reductive semisimple algebra $L'$, we provide the index $\gamma=\gamma_{L,L'}$ of $L'$ in $L$.

\subsection{Rank 2}\ 
The Lie algebras $B_2$ and $C_2$ and their Weyl groups are isomorphic. A practical difference between the two cases is in our numbering convention of simple roots (Fig.~1). In this subsection we work with $B_2$.

The branching rules for the case $B_2\supset A_1\times U_1$ are determined by the projection matrix 
$\left( \begin{smallmatrix}2&1\\ \c&1\end{smallmatrix}\right).$
In particular, for the two lowest orbits each containing 4 points, we have
$(1,0) \supset (2)(0)+(0)(2)+(0)(-2)$ and $(0,1) \supset (1)(1)+(1)(-1)$. More generally:
\begin{equation*}
\begin{aligned}          
(a,0) &\supset (2a)(0)+(0)(2a)+(0)(-2a)\,,\\
(0,b) &\supset (b)(b)+(b)(-b)\,,\\
(a,b) &\supset (2a{+}b)(b)+(2a{+}b)(-b)+(b)(2a{+}b)+(b)(-2a{-}b)\,.
\end{aligned} \quad 
a,b \in \R^{>0}
\end{equation*}
Note that the corresponding branching rules for irreducible representations are different in all cases but $(0,1)$.

The maximal subalgebra $A_1 \subset B_2$ is different than the subalgebra $A_1$ in $A_1 \times U_1 \subset B_2$. Indeed, the projection matrix for the case $B_2\supset A_1$ is $\left( \begin{smallmatrix}4&3 \end{smallmatrix}\right)$ and yields the following branching rules for the orbits:
\begin{equation*}
\begin{aligned}          
(a,0) &\supset (4a)+(2a)\,,\\
(0,b) &\supset (3b)+(b)\,,\\
(a,b) &\supset (4a{+}3b)+(2a{+}3b)+(4a{+}b)+(|2a{-}b|)\,, \\
(a,2a) &\supset (10a)+(8a)+(6a)+2(0)\,.
\end{aligned} \quad 
a,b \in \R^{>0}
\end{equation*}
The index of $A_1$ in $B_2$ is $\gamma=\gamma_{B_2,A_1}=1/5$.

For the $B_2\supset 2A_1$ case, the projection matrix $\left( \begin{smallmatrix} 1 & 1 \\ 1 & \c 
          \end{smallmatrix}\right)$
applied to the three non zero orbits gives the following branching rules:
\begin{equation*}
\begin{aligned}          
(a,0) &\supset (a)(a)\,,\\
(0,b) &\supset (b)(0)+(0)(b)\,,\\
(a,b) &\supset (a{+}b)(a)+(a)(a{+}b)\,.
\end{aligned} \quad 
a,b \in \R^{>0}
\end{equation*}
The index of $2A_1$ in $B_2$ is $\gamma=\gamma_{B_2,2A_1}=1$.

Note that in all cases the branching rules hold even if $a$ and $b$ are not integers.

\subsection{Rank 3}\
There are four cases to consider. The first one is a special case of the general case of subsection 3.8.1, except that it implies a renumbering of simple roots $C_2 \rightarrow B_2$ and a corresponding rearrangement of the projection matrix. 
\begin{alignat*}{2}
B_3\supset C_2\times U_1&:
   \left(\begin{smallmatrix} \c & 2 & 1 \\ 1 & \c & \c \\ \c & \c & 1
         \end{smallmatrix}\right)\,,\quad
&B_3\supset A_3&:
   \left( \begin{smallmatrix} \c & 1 & \c \\ 1 & \c & \c \\ \c & 1 & 1
          \end{smallmatrix}\right)\,,\quad \\
B_3\supset G_2&:
   \left( \begin{smallmatrix} \c & 1 & \c \\ 1 & \c & 1 
          \end{smallmatrix}\right)\,,\quad
&B_3\supset 3A_1&:
   \left(\begin{smallmatrix} 1 & 1 & \c \\ 1 & 1 & 1 \\ \c & 2 & 1
         \end{smallmatrix}\right)\,.
\end{alignat*}
As an example, we give the branching rules for the orbits of $B_3$ of size 6, 12, 8 and 48 respectively. We also give the index $\gamma=\gamma_{L,L'}$ whenever $L'$ is semisimple.
\begin{alignat*}{2}
B_3\supset &~C_2\times U_1: \\
 & (a,0,0)\supset (0,a)(0)+(0,0)(2a)+(0,0)(-2a)\,, \\
 & (0,b,0)\supset (2b,0)(0)+(0,b)(2b)+(0,b)(-2b) \,, \\
 & (0,0,c)\supset (c,0)(c)+(c,0)(-c)\,, \\
 & (a,b,c)\supset (2b{+}c,a)(c)+(2b{+}c,a)(-c)+(c,a{+}b)(2b{+}c)+(c,a{+}b)(-2b{-}c) \\
 & \qquad\qquad  +(c,b)(2a{+}2b{+}c)+(c,b)(-2a{-}2b{-}c)\,, \\
B_3\supset &~A_3: \\
 & (a,0,0)\supset (0,a,0)\,, \\
 & (0,b,0)\supset (b,0,b)\,, \\
 & (0,0,c)\supset (0,0,c)+(c,0,0)\,, \\
 & (a,b,c)\supset (b,a,b{+}c)+(b{+}c,a,b)\,, \\
 &\gamma=1\,,\\
B_3\supset &~G_2: \\
 & (a,0,0)\supset (0,a)\,, \\
 & (0,b,0)\supset (b,0)+(0,b)\,, \\
 & (0,0,c)\supset (0,c)+2(0,0)\,, \\
 & (a,a,a)\supset (a,2a)+2(2a,0)+(a,a)+2(a,0)\,, \\
 & (a,b,a)\supset (b,2a)+2(a{+}b,0)+(a,b)+ 
 \begin{cases}
 (a,b{-}a)\quad &\text{if}~a<b\\
 (b,a{-}b)\quad &\text{if}~a>b\\
  \end{cases} \,, \\
 & (a,a,c)\supset (a,a{+}c)+(a,c)+2(a,0)+ 
 \begin{cases}
 (2a,c{-}a)\quad &\text{if}~a<c\\
 (a{+}c,a{-}c)\quad &\text{if}~a>c\\
  \end{cases} \,, \\
 & (a,b,c)\supset (b,a{+}c) +
 \begin{cases}
 (a{+}b,c{-}a)\quad &\text{if}~a<c\\
 (b{+}c,a{-}c)\quad &\text{if}~a>c\\
  \end{cases} \quad +
 \begin{cases}
 (a,b{-}a)\quad &\text{if}~a<b\\
 (b,a{-}b)\quad &\text{if}~a>b\\
  \end{cases}
  \\
& \qquad\qquad +\begin{cases} 
(a,b{+}c{-}a)\quad &\text{if}~a<b{+}c\\
(b{+}c,a{-}b{-}c)\quad &\text{if}~a>b{+}c\\
  \end{cases} \,, \\
 &\gamma=3/2\,,
\end{alignat*}
\begin{alignat*}{2}
B_3\supset &~3A_1: \\
 & (a,0,0)\supset (a)(a)(0)+(0)(0)(2a)\,, \\
 & (0,b,0)\supset (b)(b)(2b)+(2b)(0)(0)+(0)(2b)(0)\,, \\
 & (0,0,c)\supset (0)(c)(c)+(c)(0)(c)\,, \\
 & (a,b,c)\supset (a{+}b)(a{+}b{+}c)(2b{+}c)+(b)(b{+}c)(2a{+}2b{+}c)+(a)(a{+}2b{+}c)(c) \\
 &  \qquad\qquad+(a{+}b{+}c)(a{+}b)(2b{+}c)+(b{+}c)(b)(2a{+}2b{+}c)+(a{+}2b{+}c)(a)(c)\,, \\
 &\gamma=3/4\,,
\end{alignat*}
where $a,b,c\in\R^{>0}$.

$B_3$ does not contain the principal 3-dimensional subalgebra $A_1$ as a maximal subalgebra. The corresponding $A_1$ occurs in the exceptional chain $B_3 \supset G_2 \supset A_1$. Hence the reduction from $B_3 \supset A_1$ has to be done by multiplying the projection matrices for $B_3 \supset G_2$ and $G_2 \supset A_1$, namely :
\begin{equation}
\left( \begin{smallmatrix} 10 & 6 \end{smallmatrix}\right)\left( \begin{smallmatrix} \c & 1 & \c \\ 1 & \c & 1 
          \end{smallmatrix}\right)
= \left( \begin{smallmatrix} 6 & 10 & 6 \end{smallmatrix}\right) \,. \notag         
\end{equation}          
The projection matrix obtained is the same as the one we would get from the matrix \eqref{BnA1} of the subsection 3.8.8 with $n=3$.

\subsection{Rank 4}\
There are six cases to consider. The first two are special cases of the general rank of $B_n$ in subsections 3.8.1 and 3.8.2 respectively. The next two, $B_4\supset A_3\times A_1$ and $B_4\supset C_2\times 2A_1$, are also special cases of subsections 3.8.3 and 3.8.4 respectively, except that they imply a renumbering of simple roots, $A_3 \rightarrow D_3$ and $C_2 \rightarrow B_2$, and a corresponding rearrangement of the projection matrices. The projection matrix and one example of branching rule in the case of the principal 3-dimensional subalgebra are given for the general rank, $B_n\supset A_1$, in subsection 3.8.8. 
\begin{alignat*}{3}
B_4\supset B_3\times U_1&: 
   \left(\begin{smallmatrix} 1 & \c & \c & \c \\ \c & 1 & \c & \c \\
   			\c & \c & 2 & 1 \\ \c & \c & \c & 1 
         \end{smallmatrix}\right)\,, \quad
&B_4\supset D_4&:  
   \left(\begin{smallmatrix} 1 & \c & \c & \c \\ \c & 1 & \c & \c \\
                             \c & \c & 1 & \c \\ \c & \c & 1 & 1 
         \end{smallmatrix}\right)\,,\quad
&B_4\supset A_3\times A_1&: 
   \left(\begin{smallmatrix}\c & 1 & 1 & \c \\ 
                             1 & \c & \c & \c\\ \c & 1 & 1 & 1 \\  \c & \c & 2 & 1 
         \end{smallmatrix}\right)\,, \quad \\
B_4\supset C_2\times 2A_1&:
   \left(\begin{smallmatrix} \c & \c & 2 & 1\\ 1 & 1 & \c & \c \\ 
                               \c & 1 & 1 & 1 \\ \c & 1 & 1 & \c                             
         \end{smallmatrix}\right)\,,\quad 
&B_4\supset A_1&:
   \left(\begin{smallmatrix} 8 & 14 & 18 & 10 
         \end{smallmatrix}\right)\,,\quad 
&B_4\supset 2A_1&: 
   \left(\begin{smallmatrix} 2 & 2 & 4 & 1 \\ 2 & 4 & 4 & 3 
         \end{smallmatrix}\right)\,.\quad 
\end{alignat*}
We bring here some examples of branching rules for the $B_4\supset A_1$ and $B_4\supset 2A_1$ cases, for orbits of size 8, 24, 32 and 16 respectively, together with their corresponding indices $\gamma$.
\begin{alignat*}{1}
B_4\supset &~A_1:  \\
 & (a,0,0,0)\supset (8a)+(6a)+(4a)+(2a)\,, \\
 & (0,b,0,0)\supset (14b)+(12b)+2(10b)+(8b)+2(6b)+2(4b)+3(2b)\,, \\
 & (0,0,c,0)\supset (18c)+(16c)+(14c)+2(12c)+2(10c)+(8c)+2(6c) \\
 & \qquad\qquad\quad +2(4c)+2(2c)+4(0)\,, \\
 & (0,0,0,d)\supset (10d)+(8d)+(6d)+2(4d)+2(2d)+2(0) \,, \\
 &\gamma=1/15\,,
\end{alignat*}
\begin{alignat*}{1}
B_4\supset &~2A_1: \\
 & (a,0,0,0)\supset (2a)(2a)+(0)(2a)+(2a)(0)\,, \\
 & (0,b,0,0)\supset (2b)(4b)+(4b)(2b)+(2b)(2b)+(0)(4b)+(4b)(0) \\
 & \qquad\qquad\quad +2(0)(2b)+2(2b)(0)\,, \\
 & (0,0,c,0)\supset (4c)(4c)+(0)(6c)+(6c)(0)+(2c)(4c)+(4c)(2c) \\
 & \qquad\qquad\quad +2(0)(4c)+2(4c)(0)+(0)(2c)+(2c)(0)+4(0)(0)\,, \\
 & (0,0,0,d)\supset (d)(3d)+(3d)(d)+2(d)(d) \,, \\
 &\gamma=1/3 \,,
\end{alignat*}
where $a,b,c,d\in\R^{>0}$.

\pagebreak
For cases of rank 5 to 8, we give the projection matrices which are all, except for the $B_7\supset A_3$ and $B_7\supset C_2\times A_1$ ones, special cases of the general rank section. We refrain to give the branching rules here, except for the $B_7\supset A_3$ and $B_7\supset C_2\times A_1$ cases, since they can easily be found in the general rank section, with maximally a minor renumbering of simple roots ($A_3 \rightarrow D_3$ and $C_2 \rightarrow B_2$). 

\subsection{Rank 5}\
We give the projection matrices for the six cases to consider. Examples of branching rules can be found in the corresponding subsections of the general rank section 3.8.
\begin{alignat*}{2}
B_5\supset B_4\times U_1 &:\quad
   \left(\begin{smallmatrix} 1 & \c & \c & \c & \c \\
                             \c & 1 & \c & \c & \c \\ 
                             \c & \c & 1 & \c & \c \\ 
                             \c & \c & \c & 2 & 1 \\ 
                             \c & \c & \c & \c & 1
          \end{smallmatrix}\right)\,,\qquad
&B_5\supset D_5 &:\quad
   \left(\begin{smallmatrix} 1 & \c & \c & \c & \c \\
                             \c & 1 & \c & \c & \c \\ 
                             \c & \c & 1 & \c &\c \\ 
                             \c & \c & \c & 1 & \c \\ 
                             \c & \c & \c & 1 & 1
          \end{smallmatrix}\right)\,,\qquad \\
B_5\supset B_3\times 2A_1 &:\quad
   \left(\begin{smallmatrix} 
                              1 & \c & \c & \c & \c \\
                             \c &  1 &  1 & \c & \c \\ 
                             \c & \c & \c &  2 &  1 \\
                             \c & \c &  1 &  1 & \c \\ 
                             \c & \c &  1 &  1 &  1  
          \end{smallmatrix}\right)\,,\qquad  
&B_5\supset D_4\times A_1 &:\quad
   \left(\begin{smallmatrix} 
                             1 & \c & \c & \c & \c \\ 
                             \c & 1 & \c & \c & \c \\ 
                             \c & \c & 1 & 1 & \c \\ 
                             \c & \c & 1 & 1 & 1 \\
                             \c & \c & \c & 2 & 1 
          \end{smallmatrix}\right)\,,\qquad  \\
B_5\supset A_3\times C_2 &:\quad  
   \left(\begin{smallmatrix} \c & \c & 1 & 1 & \c \\
                             1 & 1 & \c & \c & \c \\ 
                             \c & \c & 1 & 1 & \c \\ 
                             \c & \c & \c & 2 & 1 \\ 
                             \c & 1 & 1 & \c & \c
          \end{smallmatrix}\right)\,,\qquad 
&B_5\supset A_1 &:\quad
   \left(\begin{smallmatrix} 10 & 18 &  24 &  28 & 15 
         \end{smallmatrix}\right)\,. \qquad
\end{alignat*}

\subsection{Rank 6}\
We give the projection matrices for the seven cases to consider. Examples of branching rules can be found in the corresponding subsections of the general rank section 3.8.
\begin{alignat*}{2}
B_6\supset B_5\times U_1 &:
   \left(\begin{smallmatrix} 1 & \c & \c & \c & \c & \c \\
                             \c & 1 & \c & \c & \c & \c \\ 
                             \c & \c & 1 & \c & \c & \c \\ 
                             \c & \c & \c & 1 & \c & \c \\ 
                             \c & \c & \c & \c & 2 & 1 \\ 
                             \c & \c & \c & \c & \c & 1
          \end{smallmatrix}\right)\,,\qquad 
&B_6\supset D_6 &:\left(
\begin{smallmatrix} 1 & \c & \c & \c & \c & \c \\
                   \c &  1 & \c & \c & \c & \c \\ 
                   \c & \c &  1 & \c & \c & \c \\
                   \c & \c & \c &  1 & \c & \c \\ 
                   \c & \c & \c & \c &  1 & \c \\ 
                   \c & \c & \c & \c &  1 & 1
\end{smallmatrix}\right)\,,\qquad \\
B_6\supset B_4\times 2A_1 &:\left(
\begin{smallmatrix}
                   1 & \c & \c & \c & \c & \c \\
                   \c & 1 & \c & \c & \c & \c \\ 
                   \c & \c & 1 & 1 & \c & \c \\ 
                   \c & \c & \c & \c & 2 & 1 \\
                   \c & \c & \c & 1 & 1 & \c \\
                   \c & \c & \c & 1 & 1 & 1 
\end{smallmatrix}\right)\,,\qquad 
&B_6\supset D_5\times A_1 &: \left(
\begin{smallmatrix}
                   1 & \c & \c & \c & \c & \c \\ 
                   \c & 1 & \c & \c & \c & \c \\
                   \c & \c & 1 & \c & \c & \c \\ 
                   \c & \c & \c & 1 & 1 & \c \\ 
                   \c & \c & \c & 1 & 1 & 1 \\
                   \c & \c & \c & \c & 2 & 1 
\end{smallmatrix}\right)\,,\qquad \\
B_6\supset B_3\times A_3 &:\left(
\begin{smallmatrix}
                   1 & 1 & \c & \c & \c & \c \\ 
                   \c & \c & 1 & 1 & \c & \c \\ 
                   \c & \c & \c & \c & 2 & 1 \\
                   \c & \c & \c & 1 & 1 & \c \\
                   \c & 1 & 1 & \c & \c & \c \\ 
                   \c & \c & \c & 1 & 1 & 1 
\end{smallmatrix}\right)\,,\qquad 
&B_6\supset D_4\times C_2 &: \left(
\begin{smallmatrix}
                   1 & \c & \c & \c & \c & \c \\
                   \c & 1 & 1 & \c & \c & \c \\ 
                   \c & \c & \c & 1 & 1 & \c \\ 
                   \c & \c & \c & 1 & 1 & 1 \\
                   \c & \c & \c & \c & 2 & 1 \\
                   \c & \c & 1 & 1 & \c & \c 
\end{smallmatrix}\right)\,,\qquad \\
B_6\supset A_1 &: \left(
\begin{smallmatrix} 12 & 22 & 30 & 36 & 40 & 21 
\end{smallmatrix}\right)\,.\qquad 
\end{alignat*}

\subsection{Rank 7}\
We give the projection matrices of the ten cases to consider. Examples of branching rules for the first eight cases can be found in the corresponding subsections of the general rank section 3.8.
\begin{alignat*}{2}
B_7\supset B_6\times U_1 &:
   \left(\begin{smallmatrix} 1 & \c & \c & \c & \c & \c & \c \\
                             \c & 1 & \c & \c & \c & \c & \c \\ 
                             \c & \c & 1 & \c & \c & \c & \c \\ 
                             \c & \c & \c & 1 & \c & \c & \c \\ 
                             \c & \c & \c & \c & 1 & \c & \c \\
                             \c & \c & \c & \c & \c & 2 & 1 \\ 
                             \c & \c & \c & \c & \c & \c & 1
          \end{smallmatrix}\right)\,,\qquad 
&B_7\supset D_7 &:\left(
\begin{smallmatrix} 1 & \c & \c & \c & \c & \c & \c \\
                   \c &  1 & \c & \c & \c & \c & \c \\
                   \c & \c & 1 & \c & \c & \c & \c \\ 
                   \c & \c & \c &  1 & \c & \c & \c \\ 
                   \c & \c & \c & \c &  1 & \c & \c \\ 
                   \c & \c & \c & \c & \c &  1 & \c \\ 
                   \c & \c & \c & \c & \c &  1 & 1
\end{smallmatrix}\right)\,,\qquad \\
B_7\supset D_6\times A_1 &: \left(
\begin{smallmatrix}  
                   1 &  \c & \c & \c & \c & \c & \c \\
                   \c & 1 & \c & \c & \c & \c & \c \\ 
                   \c & \c & 1 &  \c & \c & \c & \c \\ 
                   \c & \c & \c & 1 &  \c & \c & \c \\ 
                   \c & \c & \c & \c & 1 & 1 & \c \\ 
                   \c & \c & \c & \c & 1 & 1 & 1 \\
                   \c & \c & \c & \c & \c & 2 & 1
\end{smallmatrix}\right)\,,\qquad 
&B_7\supset B_5\times 2A_1 &: \left(
\begin{smallmatrix} 
                   1 & \c & \c & \c & \c & \c & \c \\ 
                   \c & 1 & \c &  \c & \c & \c & \c \\ 
                   \c & \c & 1 & \c &  \c & \c & \c \\ 
                   \c & \c & \c & 1 & 1 & \c & \c \\ 
                   \c & \c & \c & \c & \c & 2 & 1 \\
                   \c & \c & \c & \c & 1 & 1 & \c \\
                   \c &  \c & \c & \c & 1 & 1 & 1 
\end{smallmatrix}\right)\,,\qquad \\ 
B_7\supset D_5\times C_2 &: \left(
\begin{smallmatrix}
                   1 & \c & \c & \c & \c & \c & \c \\ 
                   \c & 1 & \c &  \c & \c & \c & \c \\ 
                   \c & \c & 1 & 1 &  \c & \c & \c \\ 
                   \c & \c & \c & \c & 1 & 1 & \c \\ 
                   \c & \c & \c & \c & 1 & 1 & 1 \\
                    \c & \c & \c & \c & \c & 2 & 1 \\
                   \c &  \c & \c & 1 & 1 & \c & \c 
\end{smallmatrix}\right)\,,\qquad 
&B_7\supset B_4\times A_3 &: \left(
\begin{smallmatrix} 
                   1 & \c & \c &  \c & \c & \c & \c \\ 
                   \c & 1 & 1 & \c &  \c & \c & \c \\ 
                   \c & \c & \c & 1 & 1 & \c & \c \\ 
                   \c & \c & \c & \c & \c & 2 & 1 \\
                   \c & \c & \c & \c & 1 & 1 & \c \\
                   \c &  \c & 1 & 1 & \c & \c & \c \\
                   \c & \c & \c & \c & 1 & 1 & 1 
\end{smallmatrix}\right)\,,\qquad \\
B_7\supset D_4\times B_3 &: \left(
\begin{smallmatrix} 
                   1 & 1 & \c &  \c & \c & \c & \c \\ 
                   \c & \c & 1 & 1 &  \c & \c & \c \\ 
                   \c & \c & \c & \c & 1 & 1 & \c \\ 
                   \c & \c & \c & \c & 1 & 1 & 1 \\
                   \c & 1 & 1 & \c & \c & \c & \c \\
                   \c &  \c & \c & 1 & 1 & \c & \c \\
                   \c & \c & \c & \c & \c & 2 & 1 
\end{smallmatrix}\right)\,,\qquad
&B_7\supset A_1 &: \left(
\begin{smallmatrix} 14 & 26 & 36 & 44 & 50 & 54 & 28
\end{smallmatrix}\right)\,,\qquad \\
B_7\supset A_3 &: \left(
\begin{smallmatrix} 1 & \c & 1 & 1 & \c & 2 & 1 \\
                   \c &  1 & 2 & 1 & 3 & 2 & 1 \\
                   1 & 2 & 1 & 3 & 2 & 2 & 1   
\end{smallmatrix}\right)\,,\qquad 
&B_7\supset C_2\times A_1 &: \left(
\begin{smallmatrix}
                   \c &  \c & 2 & 2 & 4 & 4 & 3 \\
                   1 & 2 & 1 & 2 & 1 & 1 & \c \\
                    2 & 2 & 4 & 2 & 2 & 4 & 1 
\end{smallmatrix}\right)\,.
\end{alignat*}
We give here some examples of branching rules for the $B_7\supset A_3$ and $B_7\supset C_2\times A_1$ cases, for orbits of size 14, 84 and 128 respectively, together with their corresponding indices $\gamma$.
\begin{alignat*}{1}
B_7\supset &~A_3:\\
 & (a,0,0,0,0,0,0)\supset (a,0,a)+2(0,0,0)\,, \\
 & (0,b,0,0,0,0,0)\supset (0,b,2b)+(2b,b,0)+2(0,2b,0)+4(b,0,b)\,, \\
 & (0,0,0,0,0,0,c)\supset 2(c,c,c)+4(0,0,2c)+4(2c,0,0)+8(0,c,0)\,, \\
 &\gamma=7/12\,,\\
B_7\supset &~C_2\times A_1: \\
 & (a,0,0,0,0,0,0)\supset (0,a)(2a)+(0,a)(0)+(0,0)(2a)\,, \\
 & (0,b,0,0,0,0,0)\supset (2b,0)(4b)+2(2b,0)(2b)+3(2b,0)(0)+(0,2b)(2b)\\
 & \qquad\qquad\qquad\quad +(0,2b)(0)+(0,b)(4b)+(0,b)(2b)+2(0,b)(0)+2(0,0)(4b)\\
 & \qquad\qquad\qquad\quad +4(0,0)(2b)\,, \\
 & (0,0,0,0,0,0,c)\supset (3c,0)(c)+(c,c)(3c)+2(c,c)(c)+(c,0)(5c)+3(c,0)(3c) \\
 & \qquad\qquad\qquad\quad +5(c,0)(c)\,, \\
 &\gamma=7/16\,, 
\end{alignat*}
where $a,b,c\in\R^{>0}$.

\pagebreak
\subsection{Rank 8}\
We give the projection matrices for the nine cases to consider. Examples of branching rules can be found in the corresponding subsections of the general rank section 3.8.
\begin{alignat*}{2}
B_8\supset B_7\times U_1 &:
   \left(\begin{smallmatrix} 1 & \c & \c & \c & \c & \c & \c & \c \\
                             \c & 1 & \c & \c & \c & \c & \c & \c \\ 
                             \c & \c & 1 & \c & \c & \c & \c & \c \\ 
                             \c & \c & \c & 1 & \c & \c & \c & \c \\ 
                             \c & \c & \c & \c & 1 & \c & \c & \c \\
                             \c & \c & \c & \c & \c & 1 & \c & \c \\
                             \c & \c & \c & \c & \c & \c & 2 & 1 \\ 
                             \c & \c & \c & \c & \c & \c & \c & 1
          \end{smallmatrix}\right)\,,\qquad 
&B_8\supset D_8 &:\left(
\begin{smallmatrix} 1 & \c & \c & \c & \c & \c & \c & \c \\       
                   \c &  1 & \c & \c & \c & \c & \c & \c \\
                   \c & \c &  1 & \c & \c & \c & \c & \c \\
                   \c & \c & \c &  1 & \c & \c & \c & \c \\
                   \c & \c & \c & \c &  1 & \c & \c & \c \\ 
                   \c & \c & \c & \c & \c &  1 & \c & \c \\
                   \c & \c & \c & \c & \c & \c &  1 & \c \\
                   \c & \c & \c & \c & \c & \c &  1 & 1
\end{smallmatrix}\right)\,,\qquad \\
B_8\supset D_7\times A_1 &: \left(
\begin{smallmatrix}     
                   1 & \c & \c & \c & \c & \c & \c & \c \\
                   \c & 1 & \c & \c & \c & \c & \c & \c \\
                   \c & \c & 1 & \c & \c & \c & \c & \c \\
                   \c & \c & \c & 1 & \c & \c & \c & \c \\ 
                   \c & \c & \c & \c & 1 & \c & \c & \c \\
                   \c & \c & \c & \c & \c & 1 & 1 & \c \\
                   \c & \c & \c & \c & \c & 1 & 1 & 1 \\
                   \c & \c & \c & \c & \c & \c & 2 & 1 
\end{smallmatrix}\right)\,,\qquad 
&B_8\supset B_6\times 2A_1 &: \left(
\begin{smallmatrix} 
                   1 & \c & \c & \c & \c & \c & \c & \c \\
                   \c & 1 & \c & \c & \c & \c & \c & \c \\
                   \c & \c & 1 & \c & \c & \c & \c & \c \\ 
                   \c & \c & \c & 1 & \c & \c & \c & \c \\
                   \c & \c & \c & \c & 1 & 1 & \c & \c \\
                   \c & \c & \c & \c & \c & \c & 2 & 1 \\
                   \c & \c & \c & \c & \c & 1 & 1 & \c \\       
                   \c & \c & \c & \c & \c & 1 & 1 & 1 
\end{smallmatrix}\right)\,,\qquad \\
B_8\supset D_6\times C_2 &: \left(
\begin{smallmatrix} 
                   1 & \c & \c & \c & \c & \c & \c & \c \\
                   \c & 1 & \c & \c & \c & \c & \c & \c \\
                   \c & \c & 1 & \c & \c & \c & \c & \c \\ 
                   \c & \c & \c & 1 & 1 & \c & \c & \c \\
                   \c & \c & \c & \c & \c & 1 & 1 & \c \\
                   \c & \c & \c & \c & \c & 1 & 1 & 1 \\
                   \c & \c & \c & \c & \c & \c & 2 & 1 \\       
                   \c & \c & \c & \c & 1 & 1 & \c & \c 
\end{smallmatrix}\right)\,,\qquad 
&B_8\supset B_5\times A_3 &: \left(
\begin{smallmatrix} 
                   1 & \c & \c & \c & \c & \c & \c & \c \\
                   \c & 1 & \c & \c & \c & \c & \c & \c \\ 
                   \c & \c & 1 & 1 & \c & \c & \c & \c \\
                   \c & \c & \c & \c & 1 & 1 & \c & \c \\
                   \c & \c & \c & \c & \c & \c & 2 & 1 \\
                   \c & \c & \c & \c & \c & 1 & 1 & \c \\       
                   \c & \c & \c & 1 & 1 & \c & \c & \c \\
                   \c & \c & \c & \c & \c & 1 & 1 & 1 
\end{smallmatrix}\right)\,,\qquad \\
B_8\supset D_5\times B_3 &: \left(
\begin{smallmatrix} 
                   1 & \c & \c & \c & \c & \c & \c & \c \\
                   \c & 1 & 1 & \c & \c & \c & \c & \c \\ 
                   \c & \c & \c & 1 & 1 & \c & \c & \c \\
                   \c & \c & \c & \c & \c & 1 & 1 & \c \\
                   \c & \c & \c & \c & \c & 1 & 1 & 1 \\
                   \c & \c & 1 & 1 & \c & \c & \c & \c \\       
                   \c & \c & \c & \c & 1 & 1 & \c & \c \\
                   \c & \c & \c & \c & \c & \c & 2 & 1 
\end{smallmatrix}\right)\,,\qquad 
&B_8\supset B_4\times D_4 &: \left(
\begin{smallmatrix} 1 & 1 & \c & \c & \c & \c & \c & \c \\       
                   \c & \c & 1 & 1 & \c & \c & \c & \c \\
                   \c & \c & \c & \c & 1 & 1 & \c & \c \\
                   \c & \c & \c & \c & \c & \c & 2 & 1 \\
                   \c & 1 & 1 & \c & \c & \c & \c & \c \\ 
                   \c & \c & \c & 1 & 1 & \c & \c & \c \\
                   \c & \c & \c & \c & \c & 1 & 1 & \c \\
                   \c & \c & \c & \c & \c & 1 & 1 & 1
\end{smallmatrix}\right)\,,\qquad \\
B_8\supset A_1 &: \left(
\begin{smallmatrix} 16 & 30 & 42 & 52 & 60 & 66 & 70 & 36 
\end{smallmatrix}\right)\,.\qquad 
\end{alignat*}

\subsection{The general rank cases}\
In this section we consider infinite series of cases where the ranks of the Lie algebras take all the consecutive values starting from a lowest one. For each case, we give the corresponding projection matrix and some examples of branching rules. When the maximal reductive subalgebra of $B_n$ is semisimple, we provide also its index $\gamma$ in the Lie algebra $B_n$.

\subsubsection{$B_n\supset B_{n-1}\times U_1$, \ $(n\geq3)$}\ 
\begin{gather*}
\left(
\begin{array}{ccc}
{I_{n-2}}&\multicolumn{2}{|c}{\bf 0}\\[1ex]
\hline
\multicolumn{1}{c|}{ }&\text{{\tiny2}}&\text{{\tiny1}}\\[-1.5 ex]
\multicolumn{1}{c|}{\bf 0}&\c&\text{{\tiny1}}
\end{array}
\right)
\end{gather*}

Note that, here and everywhere below, $I_k$ denotes the $k\times k$ identity matrix, ${\bf 0}$ represents the zero matrix, and $a,b,c\in\R^{>0}$.
\begin{align*}
(a,0,0,\dots,0)  &\supset (a,0,\dots,0)(0)+(0,\dots,0)(2a)+(0,\dots,0)(-2a) \\
(0,b,0,\dots,0)  &\supset (0,b,0,\dots,0)(0)+(b,0,\dots,0)(2b)+(b,0,\dots,0)(-2b) \\
(0,0,\dots,0,c)&\supset (0,\dots,0,c)(c)+(0,\dots,0,c)(-c)
\end{align*}

Note that, here and everywhere below, in the case of $B_2$, $(0,b,0,\dots,0)$ becomes $(0,2b)$.

\pagebreak
\subsubsection{$B_{n}\supset D_{n}, \quad n\geq4$}\
\begin{gather*}
\left(
\begin{array}{ccc}
{I_{n-2}}&\multicolumn{2}{|c}{\bf 0}\\[1ex]
\hline
\multicolumn{1}{c|}{ }&\text{{\tiny1}}&\c\\[-1.5 ex]
\multicolumn{1}{c|}{\bf 0}&\text{{\tiny1}}&\text{{\tiny1}}
\end{array}
\right)
\end{gather*}

\begin{alignat*}{2}
&(a,0,0,\dots,0)  &&\supset (a,0,\dots,0) \\
&(0,b,0,\dots,0)  &&\supset (0,b,0,\dots,0) \\
&(0,0,\dots,0,c) &&\supset (0,\dots,0,c)+(0,\dots,0,c,0)\\
&\gamma=1
\end{alignat*}

\subsubsection{$B_{n}\supset D_{n-1}\times A_{1}, \quad n\geq5$}\
\begin{gather*}
\left(
\begin{array}{cccc}
{I_{n-3}}&\multicolumn{3}{|c}{\bf 0}\\[1ex]
\hline
\multicolumn{1}{c|}{ }&\text{{\tiny1}}&\text{{\tiny1}}&\c\\[-1.5 ex]
\multicolumn{1}{c|}{\bf 0}&\text{{\tiny1}}&\text{{\tiny1}}&\text{{\tiny1}}\\[-1.5 ex]
\multicolumn{1}{c|}{}&\c&\text{{\tiny2}}&\text{{\tiny1}}
\end{array}
\right)
\end{gather*}

\begin{alignat*}{2}
&(a,0,0,\dots,0)  &&\supset (a,0,\dots,0)(0)+(0,\dots,0)(2a) \\
&(0,b,0,\dots,0)  &&\supset (0,b,0,\dots,0)(0)+(b,0,\dots,0)(2b) \\
&(0,0,\dots,0,c) &&\supset (0,\dots,0,c)(c)+(0,\dots,0,c,0)(c)\\
&\gamma=n/(n+1)
\end{alignat*}

\subsubsection{$B_{n}\supset B_{n-2}\times A_{1}\times A_{1}, \quad n\geq4$}\
\begin{gather*}
\left(
\begin{array}{ccccc}
{I_{n-4}}&\multicolumn{4}{|c}{\bf 0}\\[1ex]
\hline
\multicolumn{1}{c|}{ }&\text{{\tiny1}}&\text{{\tiny1}}&\c&\c\\[-1.5 ex]
\multicolumn{1}{c|}{}&\c&\c&\text{{\tiny2}}&\text{{\tiny1}}\\[-1.5 ex]
\multicolumn{1}{c|}{\bf 0}&\c&\text{{\tiny1}}&\text{{\tiny1}}&\c\\[-1.5 ex]
\multicolumn{1}{c|}{}&\c&\text{{\tiny1}}&\text{{\tiny1}}&\text{{\tiny1}}
\end{array}
\right)
\end{gather*}

\begin{alignat*}{2}
&(a,0,0,\dots,0)  &&\supset (a,0,\dots,0)(0)(0)+(0,\dots,0)(a)(a) \\
&(0,b,0,\dots,0)  &&\supset (0,b,0,\dots,0)(0)(0)+(b,0,\dots,0)(b)(b)+(0,\dots,0)(2b)(0)\\
& &&\quad+(0,\dots,0)(0)(2b) \\
&(0,0,\dots,0,c) &&\supset (0,\dots,0,c)(c)(0)+(0,\dots,0,c)(0)(c)\\
&\gamma=1
\end{alignat*} 

\subsubsection{$B_{n}\supset B_{n-3}\times A_{3}, \quad n\geq6$}\
\begin{gather*}
\left(
\begin{array}{ccccccc}
{I_{n-6}}&\multicolumn{6}{|c}{\bf 0}\\[1ex]
\hline
\multicolumn{1}{c|}{ }&\text{{\tiny1}}&\text{{\tiny1}}&\c&\c&\c&\c\\[-1.5 ex]
\multicolumn{1}{c|}{}&\c&\c&\text{{\tiny1}}&\text{{\tiny1}}&\c&\c\\[-1.5 ex]
\multicolumn{1}{c|}{}&\c&\c&\c&\c&\text{{\tiny2}}&\text{{\tiny1}}\\[-1.5 ex]
\multicolumn{1}{c|}{\bf 0}&\c&\c&\c&\text{{\tiny1}}&\text{{\tiny1}}&\c\\[-1.5 ex]
\multicolumn{1}{c|}{}&\c&\text{{\tiny1}}&\text{{\tiny1}}&\c&\c&\c\\[-1.5 ex]
\multicolumn{1}{c|}{}&\c&\c&\c&\text{{\tiny1}}&\text{{\tiny1}}&\text{{\tiny1}}
\end{array}
\right)
\end{gather*}

\begin{alignat*}{2}
&(a,0,0,\dots,0)  &&\supset (a,0,\dots,0)(0,0,0)+(0,\dots,0)(0,a,0) \\
&(0,b,0,\dots,0)  &&\supset (0,b,0,\dots,0)(0,0,0)+(b,0,\dots,0)(0,b,0)+(0,\dots,0)(b,0,b) \\
&(0,0,\dots,0,c) &&\supset (0,\dots,0,c)(c,0,0)+(0,\dots,0,c)(0,0,c)\\
&\gamma=1
\end{alignat*}

\subsubsection{$B_{n}\supset B_{n-k}\times D_{k}, \quad n-k\geq k \geq 4$}\
\begin{gather*}
\left(
\begin{array}{ccccccccccccc}
{I_{n-2k}}&\multicolumn{12}{|c}{\bf 0}\\[1ex]
\hline
\multicolumn{1}{c|}{ }&\text{{\tiny1}}&\text{{\tiny1}}&\c&\c&\c&\c&\dots&\c&\c&\c&\c&\c\\[-1.5 ex]
\multicolumn{1}{c|}{ }&\c&\c&\text{{\tiny1}}&\text{{\tiny1}}&\c&\c&{}&\c&\c&\c&\c&\c\\[-1.5 ex]
\multicolumn{1}{c|}{ }&{}&{}&{}&{}&{}&{}&\vdots&{}&{}&{}&{}&{}\\[-1.5 ex]
\multicolumn{1}{c|}{ }&\c&\c&\c&\c&\c&\c&{}&\c&\text{{\tiny1}}&\text{{\tiny1}}&\c&\c\\[-1.5 ex]
\multicolumn{1}{c|}{ }&\c&\c&\c&\c&\c&\c&\dots&\c&\c&\c&\text{{\tiny2}}&\text{{\tiny1}}\\[-1.5 ex]
\multicolumn{1}{c|}{\bf 0}&\c&\text{{\tiny1}}&\text{{\tiny1}}&\c&\c&\c&{}&\c&\c&\c&\c&\c\\[-1.5 ex]
\multicolumn{1}{c|}{ }&\c&\c&\c&\text{{\tiny1}}&\text{{\tiny1}}&\c&{}&\c&\c&\c&\c&\c\\[-1.5 ex]
\multicolumn{1}{c|}{ }&{}&{}&{}&{}&{}&{}&\vdots&{}&{}&{}&{}&{}\\[-1.5 ex]
\multicolumn{1}{c|}{ }&\c&\c&\c&\c&\c&\c&\dots&\c&\c&\text{{\tiny1}}&\text{{\tiny1}}&\c\\[-1.5 ex]
\multicolumn{1}{c|}{ }&\c&\c&\c&\c&\c&\c&{}&\c&\c&\text{{\tiny1}}&\text{{\tiny1}}&\text{{\tiny1}}
\end{array}
\right)
\end{gather*}

\begin{alignat*}{2}
&(a,0,0,\dots,0)  &&\supset (a,0,\dots,0)(0,\dots,0)+(0,\dots,0)(a,0,\dots,0) \\
&(0,b,0,\dots,0)  &&\supset (0,b,0,\dots,0)(0,\dots,0)+(b,0,\dots,0)(b,0,\dots,0)\\
& &&\quad+(0,\dots,0)(0,b,0,\dots,0) \\
&(0,0,\dots,0,c) &&\supset (0,\dots,0,c)(0,\dots,0,c)+(0,\dots,0,c)(0,\dots,0,c,0)\\
&\gamma=1
\end{alignat*}

\subsubsection{$B_{n}\supset D_{n-k}\times B_{k}, \quad n-k> k \geq 2, \quad n-k\geq 4$}\
\begin{gather*}
\left(
\begin{array}{ccccccccccccc}
{I_{n-2k-1}}&\multicolumn{12}{|c}{\bf 0}\\[1ex]
\hline
\multicolumn{1}{c|}{ }&\text{{\tiny1}}&\text{{\tiny1}}&\c&\c&\c&\c&\dots&\c&\c&\c&\c&\c\\[-1.5 ex]
\multicolumn{1}{c|}{ }&\c&\c&\text{{\tiny1}}&\text{{\tiny1}}&\c&\c&{}&\c&\c&\c&\c&\c\\[-1.5 ex]
\multicolumn{1}{c|}{ }&{}&{}&{}&{}&{}&{}&\vdots&{}&{}&{}&{}&{}\\[-1.5 ex]
\multicolumn{1}{c|}{ }&\c&\c&\c&\c&\c&\c&{}&\c&\c&\text{{\tiny1}}&\text{{\tiny1}}&\c\\[-1.5 ex]
\multicolumn{1}{c|}{ }&\c&\c&\c&\c&\c&\c&\dots&\c&\c&\text{{\tiny1}}&\text{{\tiny1}}&\text{{\tiny1}}\\[-1.5 ex]
\multicolumn{1}{c|}{\bf 0}&\c&\text{{\tiny1}}&\text{{\tiny1}}&\c&\c&\c&{}&\c&\c&\c&\c&\c\\[-1.5 ex]
\multicolumn{1}{c|}{ }&\c&\c&\c&\text{{\tiny1}}&\text{{\tiny1}}&\c&{}&\c&\c&\c&\c&\c\\[-1.5 ex]
\multicolumn{1}{c|}{ }&{}&{}&{}&{}&{}&{}&\vdots&{}&{}&{}&{}&{}\\[-1.5 ex]
\multicolumn{1}{c|}{ }&\c&\c&\c&\c&\c&\c&\dots&\c&\text{{\tiny1}}&\text{{\tiny1}}&\c&\c\\[-1.5 ex]
\multicolumn{1}{c|}{ }&\c&\c&\c&\c&\c&\c&{}&\c&\c&\c&\text{{\tiny2}}&\text{{\tiny1}}
\end{array}
\right)
\end{gather*}

\begin{alignat*}{2}
&(a,0,0,\dots,0)  &&\supset (a,0,\dots,0)(0,\dots,0)+(0,\dots,0)(a,0,\dots,0) \\
&(0,b,0,\dots,0)  &&\supset (0,b,0,\dots,0)(0,\dots,0)+(b,0,\dots,0)(b,0,\dots,0)\\
& &&\quad+(0,\dots,0)(0,b,0,\dots,0) \\
&(0,0,\dots,0,c) &&\supset (0,\dots,0,c)(0,\dots,0,c)+(0,\dots,0,c,0)(0,\dots,0,c)\\
&\gamma=1
\end{alignat*}

\subsubsection{$B_{n}\supset A_{1}, \quad n\geq 4$}\
The projection matrix for that case is given by
\begin{gather}\label{BnA1}
\begin{gathered}
\left(
\begin{matrix} p_1 & p_2 & p_3 & \dots & p_{n-1} & p_{n}
\end{matrix}\right)\  \\   
p_{k}=k(2n-k+1), \quad 1\geq k \geq n-1; \qquad p_{n}=(n+2)(n-1)/2+1\,.
\end{gathered}
\end{gather}

We bring one example of branching rule for that case, together with the index $\gamma=\gamma_{B_n,A_1}$ :
\begin{alignat*}{1}
&(a,0,\dots,0)\supset (2na)+((2n{-}2)a)+((2n{-}4)a)+\dots+(6a)+(4a)+(2a)\,,\\
&\gamma=n/(2\displaystyle\sum\limits_{i=1}^n i^2)\,.
\end{alignat*}

\section{Reduction of orbits of the Weyl group of $C_n$}
In this section, as in the previous section, we first consider all cases of dimension up to 8. In the last subsection, 4.8, we present infinite series of selected cases. For each case of the section the projection matrix is given together with examples of the corresponding reductions/branching rules. For cases involving Weyl groups of a simple algebra $L$ and a maximal reductive semisimple algebra $L'$, we provide the index $\gamma=\gamma_{L,L'}$ of $L'$ in $L$. 

\subsection{Rank 2}\
Since the Lie algebras $B_2$ and $C_2$ and their Weyl groups are isomorphic, the projection matrices and the branching rules for the $C_2$ case can be found in subsection 3.1. A practical difference between the two cases is in our numbering convention of simple roots (Fig.~1). Hence one only needs to interchange the two columns of the projection matrices of $B_2$, and to switch the two coordinates of the orbits in the branching rules of $B_2$ to obtain the results for $C_2$. 



\subsection{Rank 3}\
There are four cases to consider. The first three are special cases of the general cases presented in the subsections 4.8.2, 4.8.3 and 4.8.5 respectively.
\begin{alignat*}{2}
C_3\supset A_2\times U_1 &:\quad
   \left(\begin{smallmatrix} 1 & 1 & \c \\
                             \c & 1 & 2 \\ 
                             1 & \c & 1
          \end{smallmatrix}\right)\,,\qquad 
&C_3\supset C_2\times A_1 &:\quad
   \left(\begin{smallmatrix} 
                             1 & \c & \c \\ 
                             \c & 1 & 1 \\
                             \c & \c & 1 
          \end{smallmatrix}\right)\,,\qquad \\
C_3\supset A_1 &:\quad
   \left(\begin{smallmatrix} 5 & 8 & 9 
         \end{smallmatrix}\right)\,,\qquad
&C_3\supset 2A_1 &:\quad
   \left(\begin{smallmatrix} 1 & \c &  1 \\ 
                             2 & 4 &  4
          \end{smallmatrix}\right)\,.  
\end{alignat*}
For all four cases, we give the branching rules for the orbits of $C_3$ of size 6, 12, 8 and 48 respectively. We also give the index $\gamma=\gamma_{L,L'}$ whenever $L'$ is semisimple.
\begin{alignat*}{1}
C_3\supset &~A_2\times U_1: \\
 & (a,0,0)\supset (a,0)(a)+(0,a)(-a)\,, \\
 & (0,b,0)\supset  (b,b)(0)+(0,b)(2b)+(b,0)(-2b)\,, \\
 & (0,0,c)\supset (0,2c)(c)+(2c,0)(-c)+(0,0)(3c)+(0,0)(-3c)\,, \\
 & (a,b,c)\supset (a{+}b,b{+}2c)(a{+}c)+(b{+}2c,a{+}b)(-a{-}c)+(b,a{+}b{+}2c)(c{-}a)\\
 & \qquad \qquad+(a{+}b{+}2c,b)(a{-}c)+(a,b)(a{+}2b{+}3c)+(b,a)(-a{-}2b{-}3c)\\
& \qquad  \qquad+(a,b{+}2c)(a{+}2b{+}c)+(b{+}2c,a)(-a{-}2b{-}c)\,, \\
C_3\supset&~ C_2\times A_1: \\
 & (a,0,0)\supset (a,0)(0)+(0,0)(a)\,, \\
 & (0,b,0)\supset  (0,b)(0)+(b,0)(b)\,, \\
 & (0,0,c)\supset (0,c)(c)\,, \\
 & (a,b,c)\supset (a,b{+}c)(c)+(a{+}b,c)(b{+}c)+(b,c)(a{+}b{+}c)\,, \\
&\gamma=1\,,\\
C_3\supset &~A_1: \\
 & (a,0,0)\supset (5a)+(3a)+(a)\,, \\
 & (0,b,0)\supset  (8b)+(6b)+2(4b)+2(2b)\,, \\
 & (0,0,c)\supset (9c)+(7c)+(3c)+(c)\,, \\
&\gamma=3/35\,,\\
\end{alignat*}
\begin{alignat*}{1}
C_3\supset &~2A_1: \\
 & (a,0,0)\supset (a)(2a)+(a)(0)\,, \\
 & (0,b,0)\supset  (0)(4b)+(2b)(2b)+(2b)(0)+2(0)(2b)\,, \\
 & (0,0,c)\supset (c)(4c)+(3c)(0)+(c)(0)\,, \\
&\gamma=3/11\,.
\end{alignat*}

For cases of rank 4 to 8, we give the projection matrices for all cases. Whenever a reduction is a special case of the general rank section, we refrain to give the branching rules and the corresponding index $\gamma$ here since they can easily be found in section 4.8. 

\subsection{Rank 4}\
We give the projection matrices of the five cases to consider. Examples of branching rules for the first four cases can be found in the corresponding subsections of the general rank section 4.8.
\begin{alignat*}{3}
C_4\supset A_3\times U_1&: 
   \left(\begin{smallmatrix} 1 & 1 & \c & \c \\ \c & \c & 1 & 2 \\ 
                             \c & 1 & 1 & \c \\ 1 & \c & 1 & \c 
         \end{smallmatrix}\right)\,, \quad
&C_4\supset C_3\times A_1&: 
   \left(\begin{smallmatrix}  1 & \c & \c & \c \\ 
                             \c & 1 & \c & \c \\ \c & \c & 1 & 1 \\ \c & \c & \c & 1 \\
         \end{smallmatrix}\right)\,,\quad 
&C_4\supset 2C_2&:
   \left(\begin{smallmatrix}
                             1 & 1 & \c & \c\\ \c & \c & 1 & 1 \\
                              \c & 1 & 1 & \c \\ \c & \c & \c & 1 
         \end{smallmatrix}\right)\,,\quad \\
C_4\supset A_1&:
   \left(\begin{smallmatrix} 7 & 12 & 15 & 16 
         \end{smallmatrix}\right)\,,\quad
&C_4\supset 3A_1&: 
   \left(\begin{smallmatrix} 1 & \c & 1 & 2 \\ 1 & 2 & 1 & 2 \\ 
                             1 & 2 & 3 & 2 
         \end{smallmatrix}\right)\,.
\end{alignat*}
We give here some examples of branching rules for the $C_4 \supset 3A_1$ case, for orbits of size 8, 24 and 16 respectively, together with the index $\gamma=\gamma_{C_4,3A_1}$.
\begin{alignat*}{1}
C_4 \supset &~3A_1: \\
 & (a,0,0,0)\supset (a)(a)(a)\,, \\
 & (0,b,0,0)\supset (0)(2b)(2b)+(2b)(0)(2b)+(2b)(2b)(0)+2(2b)(0)(0)\\
 & \qquad\qquad\quad +2(0)(2b)(0)+2(0)(0)(2b)\,, \\
 & (0,0,0,c)\supset (2c)(2c)(2c)+(0)(0)(4c)+(0)(4c)(0)+(4c)(0)(0)+2(0)(0)(0)\,, \\
&\gamma=1/3\,.
\end{alignat*}

\subsection{Rank 5}\
We give the projection matrices of the five cases to consider. Examples of branching rules for the first four cases can be found in the corresponding subsections of the general rank section 4.8.
\begin{alignat*}{2}
C_5\supset A_4\times U_1 &:\quad  
   \left(\begin{smallmatrix} 1 & 1 & \c & \c & \c \\
                              \c & \c & 1 & 1 & \c \\ 
                             \c & \c & \c & 1 & 2 \\
                             \c & 1 & 1 & \c & \c \\
                             1 & \c & 1 & \c & 1 \\
          \end{smallmatrix}\right)\,,\qquad 
&C_5\supset C_4 \times A_1 &:\quad
   \left(\begin{smallmatrix} 
                             1 & \c & \c & \c & \c \\ 
                             \c & 1 & \c & \c & \c \\ 
                             \c & \c & 1 & \c & \c \\ 
                             \c & \c & \c & 1 & 1 \\
                             \c & \c & \c & \c & 1 
          \end{smallmatrix}\right)\,,  \\
C_5\supset C_3\times C_2 &:\quad
   \left(\begin{smallmatrix} 
                             1 & \c & \c & \c & \c \\ 
                             \c & 1 & 1 & \c & \c \\ 
                             \c & \c & \c & 1 & 1 \\
                             \c & \c & 1 & 1 & \c \\
                             \c & \c & \c & \c & 1 
          \end{smallmatrix}\right)\,,\qquad  
&C_5\supset A_1 &:\quad
   \left(\begin{smallmatrix} 9 & 16 & 21 & 24 & 25 
         \end{smallmatrix}\right)\,, \\
C_5\supset C_2\times A_1 &:\quad  
   \left(\begin{smallmatrix} 
                              \c & \c & 2 & 4 & 4 \\ 
                             1 & 2 & 1 & \c & \c \\
                             1 & \c & 1 & \c & 1 
          \end{smallmatrix}\right)\,.  
\end{alignat*}
We give here some examples of branching rules for the $C_5\supset C_2\times A_1$ case, for orbits of size 10, 40 and 32 respectively, together with the index $\gamma=\gamma_{C_5,C_2\times A_1}$.
\begin{alignat*}{1}
C_5 \supset &~ C_2\times A_1:   \\
 & (a,0,0,0,0)\supset (0,a)(a)+(0,0)(a)\,, \\
 & (0,b,0,0,0)\supset (0,2b)(0)+(2b,0)(2b)+(0,b)(2b)+2(2b,0)(0)+2(0,b)(0)\\
 &\qquad\qquad\quad+2(0,0)(2b)\,, \\
 & (0,0,0,0,c)\supset (4c,0)(c)+(0,2c)(3c)+(0,2c)(c)+(0,0)(5c)+(0,0)(3c) \\
 &  \qquad\qquad\quad +2(0,0)(c)\,, \\
&\gamma=5/13\,.
\end{alignat*}

\subsection{Rank 6}\
We give the projection matrices of the seven cases to consider. Examples of branching rules for the first five cases can be found in the corresponding subsections of the general rank section 4.8.
\begin{alignat*}{2}
C_6\supset A_5\times U_1 &: 
   \left(\begin{smallmatrix} 1 & 1 & \c & \c & \c & \c \\
                              \c & \c & 1 & 1 & \c & \c \\ 
                             \c & \c & \c & \c & 1 & 2 \\
                             \c & \c & \c & 1 & 1 & \c \\
                             \c & 1 & 1 & \c & \c & \c \\
                             1 & \c & 1 & \c & 1 & \c \\
          \end{smallmatrix}\right)\,,\qquad
&C_6\supset C_5 \times A_1 &:\left(
\begin{smallmatrix} 
                   1 & \c & \c & \c & \c & \c \\ 
                   \c & 1 & \c & \c & \c & \c \\
                   \c & \c & 1 & \c & \c & \c \\ 
                   \c & \c & \c & 1 & \c & \c \\ 
                   \c & \c & \c & \c & 1 & 1 \\
                   \c & \c & \c & \c & \c & 1 
\end{smallmatrix}\right)\,,\\
C_6\supset C_4\times C_2 &: \left(
\begin{smallmatrix} 
                   1 & \c & \c & \c & \c & \c \\
                   \c & 1 & \c & \c & \c & \c \\ 
                   \c & \c & 1 & 1 & \c & \c \\ 
                   \c & \c & \c & \c & 1 & 1 \\
                   \c & \c & \c & 1 & 1 & \c \\
                   \c & \c & \c & \c & \c & 1 
\end{smallmatrix}\right)\,,\qquad 
&C_6\supset 2C_3 &:\left(
\begin{smallmatrix} 
                   1 & 1 & \c & \c & \c & \c \\ 
                   \c & \c & 1 & 1 & \c & \c \\ 
                   \c & \c & \c & \c & 1 & 1 \\
                   \c & 1 & 1 & \c & \c & \c \\
                   \c & \c & \c & 1 & 1 & \c \\ 
                   \c & \c & \c & \c & \c & 1 
\end{smallmatrix}\right)\,,\\  
C_6\supset A_1 &: \left(
\begin{smallmatrix} 11 & 20 & 27 & 32 & 35 & 36
\end{smallmatrix}\right)\,,\qquad 
&C_6\supset A_3 \times A_1 &:\left(
\begin{smallmatrix} 
                   \c & \c & 1 & 2 & 1 & 2 \\ 
                   1 & 2 & 1 & \c & \c & \c \\
                   \c & \c & 1 & 2 & 3 & 2 \\
                   1 & \c & 1 & \c & 1 & 2 
\end{smallmatrix}\right)\,,\\
C_6\supset C_2\times A_1 &: \left(
\begin{smallmatrix} 
                   1 & 2 & 1 & 2 & 1 & 2 \\ 
                   \c & \c & 1 & 1 & 2 & 1 \\
                   2 & 2 & 4 & 2 & 2 & 4 
\end{smallmatrix}\right)\,.
\end{alignat*}
We give here some examples of branching rules for the $C_6\supset A_3 \times A_1$ and $C_6\supset C_2 \times A_1$ cases, for orbits of size 12, 60 and 64 respectively, together with their corresponding indices $\gamma$.
\begin{alignat*}{1}
C_6\supset &~A_3 \times A_1:\\
 & (a,0,0,0,0,0)\supset (0,a,0)(a)\,, \\
 & (0,b,0,0,0,0)\supset (0,2b,0)(0)+(b,0,b)(2b)+2(b,0,b)(0)+3(0,0,0)(2b)\,, \\
 & (0,0,0,0,0,c)\supset (2c,0,2c)(2c)+(0,0,4c)(0)+(4c,0,0)(0)+(0,2c,0)(4c)\\
 &\qquad\qquad\qquad\quad+2(0,2c,0)(0)+(0,0,0)(6c)+3(0,0,0)(2c)\,, \\
&\gamma=1/3\,,
\end{alignat*}
\begin{alignat*}{1}
C_6\supset &~C_2\times A_1: \\
 & (a,0,0,0,0,0)\supset (a,0)(2a)+(a,0)(0)\,, \\
 & (0,b,0,0,0,0)\supset (2b,0)(2b)+(0,b)(4b)+2(0,b)(2b)+(2b,0)(0)+3(0,b)(0) \\
 &\qquad\qquad\qquad\quad+2(0,0)(4b)+4(0,0)(2b)\,,\\
 & (0,0,0,0,0,c)\supset (2c,c)(4c)+(0,3c)(0)+(2c,c)(0)+(0,c)(8c)+2(0,c)(4c) \\
 &\qquad\qquad\qquad\quad+3(0,c)(0)\,, \\
&\gamma=3/11\,.
\end{alignat*}

\subsection{Rank 7}\
We give the projection matrices of the six cases to consider. Examples of branching rules for the first five cases can be found in the corresponding subsections of the general rank section 4.8.
\begin{alignat*}{2}
C_7\supset A_6\times U_1 &: \left(
\begin{smallmatrix} 1 & 1 & \c & \c & \c & \c & \c \\
                   \c & \c & 1 & 1 & \c & \c & \c \\
                   \c & \c & \c & \c & 1 & 1 & \c \\ 
                   \c & \c & \c & \c & \c & 1 & 2 \\ 
                   \c & \c & \c & 1 & 1 & \c & \c \\ 
                   \c & 1 & 1 & \c & \c & \c & \c \\ 
                   1 & \c & 1 & \c & 1 & \c & 1 
\end{smallmatrix}\right)\,,\qquad
&C_7\supset C_6 \times A_1 &: \left(
\begin{smallmatrix} 
                   1 & \c & \c & \c & \c & \c & \c \\
                   \c & 1 & \c & \c & \c & \c & \c \\ 
                   \c & \c & 1 & \c & \c & \c & \c \\ 
                   \c & \c & \c & 1 & \c & \c & \c \\ 
                   \c & \c & \c & \c & 1 & \c & \c \\ 
                   \c & \c & \c & \c & \c & 1 & 1 \\
                   \c & \c & \c & \c & \c & \c & 1 
\end{smallmatrix}\right)\,,\qquad \\
C_7\supset C_5\times C_2 &: \left(
\begin{smallmatrix}
                   1 & \c & \c & \c & \c & \c & \c \\ 
                   \c & 1 & \c & \c & \c & \c & \c \\ 
                   \c & \c & 1 & \c & \c & \c & \c \\ 
                   \c & \c & \c & 1 & 1 & \c & \c \\ 
                   \c & \c & \c & \c & \c & 1 & 1 \\
                    \c & \c & \c & \c & 1 & 1 & \c \\
                   \c & \c & \c & \c & \c & \c & 1  
\end{smallmatrix}\right)\,,\qquad
&C_7\supset C_4\times C_3 &: \left(
\begin{smallmatrix} 
                   1 & \c & \c & \c & \c & \c & \c \\ 
                   \c & 1 & 1 & \c & \c & \c & \c \\ 
                   \c & \c & \c & 1 & 1 & \c & \c \\ 
                   \c & \c & \c & \c & \c & 1 & 1 \\
                    \c & \c & 1 & 1 & \c & \c & \c \\
                   \c & \c & \c & \c & 1 & 1 & \c \\
                   \c & \c & \c & \c & \c & \c & 1 
\end{smallmatrix}\right)\,, \\ 
C_7\supset A_1 &: \left(
\begin{smallmatrix} 13 & 24 & 33 & 40 & 45 & 48 & 49
\end{smallmatrix}\right)\,,\qquad 
&C_7\supset B_3 \times A_1 &: \left(
\begin{smallmatrix} 
                   1 & 2 & 1 & \c & \c & \c & \c \\
                   \c & \c & 1 & 2 & 1 & \c & \c \\ 
                   \c & \c & \c & \c & 2 & 4 & 4 \\
                   1 & \c & 1 & \c & 1 & \c & 1 
\end{smallmatrix}\right)\,.
\end{alignat*}
We give here some examples of branching rules for the $C_7\supset B_3 \times A_1$ case, for orbits of size 14, 84 and 128 respectively, together with the index $\gamma=\gamma_{C_7,B_3\times A_1}$.
\begin{alignat*}{2}
C_7\supset &~B_3 \times A_1 : \\
 & (a,0,0,0,0,0,0)\supset (a,0,0)(a)+(0,0,0)(a)\,, \\
 & (0,b,0,0,0,0,0)\supset (2b,0,0)(0)+(0,b,0)(2b)+2(0,b,0)(0)+(b,0,0)(2b)\\
 & \qquad \qquad \qquad \qquad+2(b,0,0)(0)+3(0,0,0)(2b)\,, \\
 & (0,0,0,0,0,0,c)\supset (0,0,4c)(c)+(0,2c,0)(3c)+(0,2c,0)(c)+(2c,0,0)(5c)\\
 & \qquad \qquad \qquad \qquad+(2c,0,0)(3c)+2(2c,0,0)(c)+(0,0,0)(7c)+(0,0,0)(5c)\\
 & \qquad \qquad \qquad \qquad+3(0,0,0)(3c)+3(0,0,0)(c)\,, \\
&\gamma=7/19\,.
\end{alignat*}

\subsection{Rank 8}\
We give the projection matrices of the eight cases to consider. Examples of branching rules for the first six cases can be found in the corresponding subsections of the general rank section 4.8.
\begin{alignat*}{2}
C_8\supset A_7\times U_1 &: \left(
\begin{smallmatrix} 1 & 1 & \c & \c & \c & \c & \c & \c \\       
                   \c & \c & 1 & 1 & \c & \c & \c & \c \\
                   \c & \c & \c & \c & 1 & 1 & \c & \c \\
                   \c & \c & \c & \c & \c & \c & 1 & 2 \\
                   \c & \c & \c & \c & \c & 1 & 1 & \c \\ 
                   \c & \c & \c & 1 & 1 & \c & \c & \c \\
                   \c & 1 & 1 & \c & \c & \c & \c & \c \\
                   1 & \c & 1 & \c & 1 & \c & 1 & \c
\end{smallmatrix}\right)\,,\qquad
&C_8\supset C_7\times A_1 &: \left(
\begin{smallmatrix}       
                   1 & \c & \c & \c & \c & \c & \c & \c \\
                   \c & 1 & \c & \c & \c & \c & \c & \c \\
                   \c & \c & 1 & \c & \c & \c & \c & \c \\
                   \c & \c & \c & 1 & \c & \c & \c & \c \\ 
                   \c & \c & \c & \c & 1 & \c & \c & \c \\
                   \c & \c & \c & \c & \c & 1 & \c & \c \\
                   \c & \c & \c & \c & \c & \c & 1 & 1 \\
                   \c & \c & \c & \c & \c & \c & \c & 1 
\end{smallmatrix}\right)\,,\\
C_8\supset C_6\times C_2 &: \left(
\begin{smallmatrix} 
                   1 & \c & \c & \c & \c & \c & \c & \c \\
                   \c & 1 & \c & \c & \c & \c & \c & \c \\
                   \c & \c & 1 & \c & \c & \c & \c & \c \\ 
                   \c & \c & \c & 1 & \c & \c & \c & \c \\
                   \c & \c & \c & \c & 1 & 1 & \c & \c \\
                   \c & \c & \c & \c & \c & \c & 1 & 1 \\
                   \c & \c & \c & \c & \c & 1 & 1 & \c \\       
                   \c & \c & \c & \c & \c & \c & \c & 1 
\end{smallmatrix}\right)\,,\qquad 
&C_8\supset C_5\times C_3 &: \left(
\begin{smallmatrix} 
                   1 & \c & \c & \c & \c & \c & \c & \c \\
                   \c & 1 & \c & \c & \c & \c & \c & \c \\ 
                   \c & \c & 1 & 1 & \c & \c & \c & \c \\
                   \c & \c & \c & \c & 1 & 1 & \c & \c \\
                   \c & \c & \c & \c & \c & \c & 1 & 1 \\
                   \c & \c & \c & 1 & 1 & \c & \c & \c \\       
                   \c & \c & \c & \c & \c & 1 & 1 & \c \\
                   \c & \c & \c & \c & \c & \c & \c & 1 
\end{smallmatrix}\right)\,,\\
C_8\supset 2C_4 &: \left(
\begin{smallmatrix} 
                   1 & 1 & \c & \c & \c & \c & \c & \c \\ 
                   \c & \c & 1 & 1 & \c & \c & \c & \c \\
                   \c & \c & \c & \c & 1 & 1 & \c & \c \\
                   \c & \c & \c & \c & \c & \c & 1 & 1 \\
                   \c & 1 & 1 & \c & \c & \c & \c & \c \\       
                   \c & \c & \c & 1 & 1 & \c & \c & \c \\
                   \c & \c & \c & \c & \c & 1 & 1 & \c \\
                   \c & \c & \c & \c & \c & \c & \c & 1 
\end{smallmatrix}\right)\,,\qquad 
&C_8\supset A_1 &: \left(
\begin{smallmatrix} 15 & 28 & 39 & 48 & 55 & 60 & 63 & 64
\end{smallmatrix}\right)\,,\\
C_8\supset D_4\times A_1 &: \left(
\begin{smallmatrix}        
                   1 & 2 & 1 & \c & \c & \c & \c & \c \\
                   \c & \c & 1 & 2 & 1 & \c & \c & \c \\
                   \c & \c & \c & \c & 1 & 2 & 1 & 2 \\
                   \c & \c & \c & \c & 1 & 2 & 3 & 2 \\
                   1 & \c & 1 & \c & 1 & \c & 1 & 2 
\end{smallmatrix}\right)\,, \qquad
&C_8\supset C_2 &: \left(
\begin{smallmatrix} 1 & 4 & 3 & 4 & 5 & 8 & 7 & 6 \\       
                   1 & \c & 2 & 2 & 2 & \c & 1 & 2
\end{smallmatrix}\right)\,. 
\end{alignat*}
We give here some examples of branching rules for the $C_8\supset D_4 \times A_1$ and $C_8\supset C_2$ cases, for orbits of size 16, 112 and 256 respectively, together with their corresponding indices $\gamma$.
\begin{alignat*}{1}
C_8\supset &~D_4\times A_1 : \\
 & (a,0,0,0,0,0,0,0)\supset (a,0,0,0)(a)\,, \\
 & (0,b,0,0,0,0,0,0)\supset (2b,0,0,0)(0)+(0,b,0,0)(2b)+2(0,b,0,0)(0)\\
& \qquad \qquad \qquad \qquad +4(0,0,0,0)(2b)\,, \\
 & (0,0,0,0,0,0,0,c)\supset (0,0,2c,2c)(2c)+(0,0,0,4c)(0)+(0,0,4c,0)(0)\\
& \qquad \qquad \qquad \qquad +(0,2c,0,0)(4c)+2(0,2c,0,0)(0)+(2c,0,0,0)(6c)\\
& \qquad \qquad \qquad \qquad +3(2c,0,0,0)(2c)+(0,0,0,0)(8c)+4(0,0,0,0)(4c)\\
& \qquad \qquad \qquad \qquad +6(0,0,0,0)(0)\,, \\
&\gamma=1/3\\
C_8\supset &~C_2 : \\
 & (a,0,0,0,0,0,0,0)\supset (a,a)+2(a,0)\,, \\
 & (0,b,0,0,0,0,0,0)\supset (4b,0)+(0,3b)+3(2b,b)+6(2b,0)+4(0,2b)+9(0,b)\\
& \qquad \qquad \qquad \qquad +4(0,0)\,, \\
 & (0,0,0,0,0,0,0,c)\supset (6c,2c)+2(8c,0)+3(4c,2c)+2(2c,4c)+4(6c,0)\\ 
& \qquad \qquad \qquad \qquad+(0,6c)+6(2c,2c)+6(4c,0)+5(0,4c)+10(2c,0)\\ 
& \qquad \qquad \qquad \qquad+9(0,2c)+12(0,0)\,, \\
&\gamma=1/3\,.
\end{alignat*}

\subsection{The general rank cases}\
In this section, we consider infinite series of cases where the ranks of the Lie algebras take all the consecutive values starting from a lowest one. For each case, we give the corresponding projection matrix and some examples of branching rules. When the maximal reductive subalgebra of $C_n$ is semisimple, we also provide  its index $\gamma$ in the Lie algebra $C_n$.

\subsubsection{$C_{2n}\supset A_{2n-1}\times U_{1}, \quad n\geq1$}

\begin{gather*}
\left(
\begin{array}{ccccccccccccc}
\text{{\tiny1}}&\text{{\tiny1}}&\c&\c&\c&\c&\dots&\c&\c&\c&\c&\c&\c\\[-1.5 ex]
\c&\c&\text{{\tiny1}}&\text{{\tiny1}}&\c&\c&{}&\c&\c&\c&\c&\c&\c\\[-1.5 ex]
{}&{}&{}&{}&{}&{}&\vdots&{}&{}&{}&{}&{}&{}\\[-1.5 ex]
\c&\c&\c&\c&\c&\c&{}&\c&\c&\text{{\tiny1}}&\text{{\tiny1}}&\c&\c\\[-1.5 ex]
\c&\c&\c&\c&\c&\c&\dots&\c&\c&\c&\c&\text{{\tiny1}}&\text{{\tiny2}}\\[-1.5 ex]
\c&\c&\c&\c&\c&\c&{}&\c&\c&\c&\text{{\tiny1}}&\text{{\tiny1}}&\c\\[-1.5 ex]
\c&\c&\c&\c&\c&\c&{}&\c&\text{{\tiny1}}&\text{{\tiny1}}&\c&\c&\c\\[-1.5 ex]
{}&{}&{}&{}&{}&{}&\vdots&{}&{}&{}&{}&{}&{}\\[-1.5 ex]
\c&\text{{\tiny1}}&\text{{\tiny1}}&\c&\c&\c&\dots&\c&\c&\c&\c&\c&\c\\[-1.5 ex]
\text{{\tiny1}}&\c&\text{{\tiny1}}&\c&\text{{\tiny1}}&\c&{}&\text{{\tiny1}}&\c&\text{{\tiny1}}&\c&\text{{\tiny1}}&\c
\end{array}
\right)
\end{gather*}

\begin{align*}
(a,0,0,\dots,0)  &\supset (a,0,\dots,0)(a)+(0,\dots,0,a)(-a) \\
(0,b,0,\dots,0)  &\supset (b,0,\dots,0,b)(0)+(0,b,0,\dots,0)(2b)+(0,\dots,0,b,0)(-2b) \\
(0,0,\dots,0,c)  &\supset
(\underbrace {0,\dots,0}_{n-1},2c,\underbrace {0,\dots,0}_{n-1})(0)
+(\underbrace {0,\dots,0}_{n},2c,\underbrace {0,\dots,0}_{n-2})(2c) \\
& \quad +(\underbrace {0,\dots,0}_{n-2},2c,\underbrace {0,\dots,0}_{n})(-2c)
+(\underbrace {0,\dots,0}_{n+1},2c,\underbrace {0,\dots,0}_{n-3})(4c) \\
& \quad +(\underbrace {0,\dots,0}_{n-3},2c,\underbrace {0,\dots,0}_{n+1})(-4c) +...
+(0,\dots,0,2c)((2n{-}2)c) \\
& \quad +(2c,0,\dots,0)({-}(2n{-}2)c) + (0,\dots,0)(2nc)
+(0,\dots,0)(-2nc)
\end{align*}

\subsubsection{$C_{2n+1}\supset A_{2n}\times U_{1}, \quad n\geq1$}

\begin{gather*}
\left(
\begin{array}{cccccccccccccc}
\text{{\tiny1}}&\text{{\tiny1}}&\c&\c&\c&\c&\dots&\c&\c&\c&\c&\c&\c&\c\\[-1.5 ex]
\c&\c&\text{{\tiny1}}&\text{{\tiny1}}&\c&\c&{}&\c&\c&\c&\c&\c&\c&\c\\[-1.5 ex]
{}&{}&{}&{}&{}&{}&\vdots&{}&{}&{}&{}&{}&{}&{}\\[-1.5 ex]
\c&\c&\c&\c&\c&\c&{}&\c&\c&\c&\c&\text{{\tiny1}}&\text{{\tiny1}}&\c\\[-1.5 ex]
\c&\c&\c&\c&\c&\c&\dots&\c&\c&\c&\c&\c&\text{{\tiny1}}&\text{{\tiny2}}\\[-1.5 ex]
\c&\c&\c&\c&\c&\c&{}&\c&\c&\c&\text{{\tiny1}}&\text{{\tiny1}}&\c&\c\\[-1.5 ex]
\c&\c&\c&\c&\c&\c&{}&\c&\text{{\tiny1}}&\text{{\tiny1}}&\c&\c&\c&\c\\[-1.5 ex]
{}&{}&{}&{}&{}&{}&\vdots&{}&{}&{}&{}&{}&{}&{}\\[-1.5 ex]
\c&\text{{\tiny1}}&\text{{\tiny1}}&\c&\c&\c&\dots&\c&\c&\c&\c&\c&\c&\c\\[-1.5 ex]
\text{{\tiny1}}&\c&\text{{\tiny1}}&\c&\text{{\tiny1}}&\c&{}&\text{{\tiny1}}&\c&\text{{\tiny1}}&\c&\text{{\tiny1}}&\c&\text{{\tiny1}}
\end{array}
\right)
\end{gather*}

\begin{align*}
(a,0,0,\dots,0)  &\supset (a,0,\dots,0)(a)+(0,\dots,0,a)({-}a) \\
(0,b,0,\dots,0)  &\supset (b,0,\dots,0,b)(0)+(0,b,0,\dots,0)(2b)+(0,\dots,0,b,0)({-}2b) \\
(0,0,\dots,0,c)  &\supset  
(\underbrace {0,\dots,0}_{n},2c,\underbrace {0,\dots,0}_{n{-}1})(c)
+(\underbrace {0,\dots,0}_{n{-}1},2c,\underbrace {0,\dots,0}_{n})(-c) \\
& \quad +(\underbrace {0,\dots,0}_{n{+}1},2c,\underbrace {0,\dots,0}_{n{-}2})(3c)
+(\underbrace {0,\dots,0}_{n{-}2},2c,\underbrace {0,\dots,0}_{n{+}1})(-3c) \\
& \quad +...
+(0,\dots,0,2c)((2n{-}1)c)
+(2c,0,\dots,0)({-}(2n{-}1)c) \\
& \quad +(0,\dots,0)((2n{+}1)c)
+(0,\dots,0)({-}(2n{+}1)c)
\end{align*}

\subsubsection{$C_n\supset C_{n-1}\times A_1$, \ $(n\geq2)$}\ 

\begin{gather*}
\left(
\begin{array}{ccc}
{I_{n-2}}&\multicolumn{2}{|c}{\bf 0}\\[1ex]
\hline
\multicolumn{1}{c|}{ }&\text{{\tiny1}}&\text{{\tiny1}}\\[-1.5 ex]
\multicolumn{1}{c|}{\bf 0}&\c&\text{{\tiny1}}
\end{array}
\right)
\end{gather*}

\begin{alignat*}{2}
&(a,0,0,\dots,0)  &&\supset (a,0,\dots,0)(0)+(0,\dots,0)(a) \\
&(0,b,0,\dots,0)  &&\supset (0,b,0,\dots,0)(0)+(b,0,\dots,0)(b) \\
&(0,0,\dots,0,c) &&\supset (0,\dots,0,c)(c) \\
&\gamma=1
\end{alignat*}

\subsubsection{$C_{n}\supset C_{n-k}\times C_{k}, \quad n-k\geq k \geq 2$}\

\begin{gather*}
\left(
\begin{array}{ccccccccccccc}
{I_{n-2k}}&\multicolumn{12}{|c}{\bf 0}\\[1ex]
\hline
\multicolumn{1}{c|}{ }&\text{{\tiny1}}&\text{{\tiny1}}&\c&\c&\c&\c&\dots&\c&\c&\c&\c&\c\\[-1.5 ex]
\multicolumn{1}{c|}{ }&\c&\c&\text{{\tiny1}}&\text{{\tiny1}}&\c&\c&{}&\c&\c&\c&\c&\c\\[-1.5 ex]
\multicolumn{1}{c|}{ }&{}&{}&{}&{}&{}&{}&\vdots&{}&{}&{}&{}&{}\\[-1.5 ex]
\multicolumn{1}{c|}{ }&\c&\c&\c&\c&\c&\c&{}&\c&\text{{\tiny1}}&\text{{\tiny1}}&\c&\c\\[-1.5 ex]
\multicolumn{1}{c|}{ }&\c&\c&\c&\c&\c&\c&\dots&\c&\c&\c&\text{{\tiny1}}&\text{{\tiny1}}\\[-1.5 ex]
\multicolumn{1}{c|}{\bf 0}&\c&\text{{\tiny1}}&\text{{\tiny1}}&\c&\c&\c&{}&\c&\c&\c&\c&\c\\[-1.5 ex]
\multicolumn{1}{c|}{ }&\c&\c&\c&\text{{\tiny1}}&\text{{\tiny1}}&\c&{}&\c&\c&\c&\c&\c\\[-1.5 ex]
\multicolumn{1}{c|}{ }&{}&{}&{}&{}&{}&{}&\vdots&{}&{}&{}&{}&{}\\[-1.5 ex]
\multicolumn{1}{c|}{ }&\c&\c&\c&\c&\c&\c&\dots&\c&\c&\text{{\tiny1}}&\text{{\tiny1}}&\c\\[-1.5 ex]
\multicolumn{1}{c|}{ }&\c&\c&\c&\c&\c&\c&{}&\c&\c&\c&\c&\text{{\tiny1}}
\end{array}
\right)
\end{gather*}

\begin{alignat*}{2}
&(a,0,0,\dots,0)  &&\supset (a,0,\dots,0)(0,\dots,0)+(0,\dots,0)(a,0,\dots,0) \\
&(0,b,0,\dots,0)  &&\supset (0,b,0,\dots,0)(0,\dots,0)+(b,0,\dots,0)(b,0,\dots,0)\\
& &&\quad+(0,\dots,0)(0,b,0,\dots,0) \\
&(0,0,\dots,0,c) &&\supset (0,\dots,0,c)(0,\dots,0,c)\\
&\gamma=1
\end{alignat*}

\subsubsection{$C_{n}\supset A_{1}, \quad n\geq2$}
The projection matrix for that case is given by
$$
\left(
\begin{matrix} p_1 & p_2 & p_3 & \dots & p_{n-1} & p_{n}
\end{matrix}\right)\     \qquad p_{k}=k(2n-k), \qquad 1\geq k \geq n .
$$
We bring one example of branching rule for that case, together with the index $\gamma=\gamma_{C_n,A_1}$ :
\begin{align*}
&(a,0,\dots,0) \supset ((2n{-}1)a)+((2n{-}3)a)+((2n{-}5)a)+\dots+(5a)+(3a)+(a)\,,\\
&\gamma=n/\displaystyle\sum\limits_{i=1}^n (2i{-}1)^2\,.
\end{align*}

\section{Reduction of orbits of the Weyl group of $D_n$}
As in the two previous sections, we first consider all cases of dimension up to 8, and we present infinite series of selected cases in 5.7. For each case,  the projection matrix is given together with examples of the corresponding branching rules. For cases involving Weyl groups of a simple algebra $L$ and a maximal reductive semisimple algebra $L'$, we provide the index $\gamma=\gamma_{L,L'}$ of $L'$ in $L$. 

\subsection{Rank 3}\
Since the Lie algebras $D_3$ and $A_3$ and their Weyl groups are isomorphic, the projection matrices and some examples of branching rules for the $D_3$ case can be found in \cite{LNP}. A practical difference between the two cases is in our numbering convention of simple roots (Fig.~1). 

For cases of rank 4 to 8, we give the projection matrices for all cases. Whenever a reduction is a special case of the general rank section, we refrain to give the branching rules and the corresponding index $\gamma$ here since they can easily be found in section 5.7, with maximally a minor renumbering of simple roots ($A_3 \rightarrow D_3$ and $C_2 \rightarrow B_2$). 

\subsection{Rank 4}\
We give the projection matrices of the five cases to consider. Examples of branching rules for the first three cases can be found in the corresponding subsections of the general rank section 5.7.
\begin{alignat}{3}
D_4\supset A_3 \times U_1&: 
   \left(\begin{smallmatrix} 1 & 1 & \c & \c \\ \c & \c & \c & 1 \\ 
                              \c & 1 & 1 & \c \\ 1 & \c & 1 & \c
         \end{smallmatrix}\right)\,, \qquad
&D_4\supset B_3&:
   \left(\begin{smallmatrix} 1 & \c & \c & \c \\ \c & 1 & \c & \c \\
                             \c & \c & 1 & 1  
         \end{smallmatrix}\right)\,,\qquad 
&D_4\supset C_2\times A_1&:
   \left(\begin{smallmatrix}\c & 2 & 1 & 1\\ 1 & \c & \c & \c \\ \c & \c & 1 & 1                               
         \end{smallmatrix}\right)\,, \qquad \notag \\
D_4\supset 4A_1&: 
   \left(\begin{smallmatrix} \c & 1 & 1 & \c \\ \c & 1 & \c & 1\\ 
                              1 & 1 & 1 & 1 \\ 1 & 1 & \c & \c
         \end{smallmatrix}\right)\,,\qquad 
&D_4\supset A_2&: 
   \left(\begin{smallmatrix} 1 & \c & 1 & 1 \\ 1 & 3 & 1 & 1
         \end{smallmatrix}\right)\,.\qquad \notag 
\end{alignat}
We give here some examples of branching rules for the $D_4\supset 4A_1$ and $D_4\supset A_2$ cases, for orbits of size 8, 24 and 8 respectively, together with their corresponding indices $\gamma$.
\begin{alignat*}{1}
 D_4\supset &~4A_1: \\
& (a,0,0,0)\supset (a)(a)(0)(0)+(0)(0)(a)(a)\,, \\
 & (0,b,0,0)\supset (b)(b)(b)(b)+(2b)(0)(0)(0)+(0)(2b)(0)(0)+(0)(0)(2b)(0) \\
 & \qquad\qquad\quad +(0)(0)(0)(2b)\,, \\
 & (0,0,0,c)\supset (0)(c)(c)(0)+(c)(0)(0)(c) \,, \\
&\gamma=1\,,\\
D_4\supset &~A_2: \\
 & (a,0,0,0)\supset (a,a)+2(0,0)\,, \\
 & (0,b,0,0)\supset (0,3b)+(3b,0)+3(b,b)\,, \\
 & (0,0,0,c)\supset  (c,c)+2(0,0)\,, \\
&\gamma=2/3\,.
\end{alignat*}

\subsection{Rank 5}\
We give the projection matrices of the seven cases to consider. Examples of branching rules for the first five cases can be found in the corresponding subsections of the general rank section 5.7.
\begin{alignat*}{2}
D_5\supset A_4\times U_1 &:\quad  
   \left(\begin{smallmatrix}  1 & 1 & \c & \c & \c \\
                             \c & \c & 1 & 1 & \c \\ 
                             \c & \c & \c & \c & 1 \\ 
                             \c & 1 & 1 & \c & \c \\
                             2 & \c & 2 & -1 & 1
          \end{smallmatrix}\right)\,,\qquad 
&D_5\supset D_4\times U_1 &:\quad
   \left(\begin{smallmatrix}  1 & \c &  \c & \c & \c \\
                             \c & 1 & \c & \c & \c \\ 
                             \c & \c & 1 & \c & \c \\ 
                             \c &  \c & 1 & 1 & 1 \\
                             \c & \c & \c & 1 & -1
          \end{smallmatrix}\right)\,,\qquad  \\
D_5\supset B_4 &:\quad
   \left(\begin{smallmatrix}  1 & \c & \c & \c & \c \\
                             \c & 1 & \c & \c & \c \\ 
                             \c & \c & 1 & \c & \c \\ 
                             \c & \c & \c & 1 & 1 
         \end{smallmatrix}\right)\,, \qquad          
&D_5\supset B_3\times A_1 &:\quad  
   \left(\begin{smallmatrix}  
                             1 & \c & \c & \c & \c \\ 
                             \c & 1 & \c & \c & \c \\ 
                             \c & \c & 2 & 1 & 1 \\
                             \c & \c & \c & 1 & 1 
          \end{smallmatrix}\right)\,,\qquad \\
D_5\supset 2C_2 &:\quad
   \left(\begin{smallmatrix}  \c & \c & 2 & 1 & 1 \\
                             1 & 1 & \c & \c & \c \\ 
                             \c & \c & \c & 1 & 1 \\ 
                             \c & 1 & 1 & \c & \c
          \end{smallmatrix}\right)\,,\qquad 
&D_5\supset A_3\times 2A_1 &:\quad
   \left(\begin{smallmatrix}  
                             \c & \c & 1 & 1 & \c \\ 
                             1 & 1 & \c & \c & \c \\ 
                             \c & \c & 1 & \c & 1\\
                             \c & 1 & 1 & 1 & 1 \\
                             \c & 1 & 1 &  \c & \c 
          \end{smallmatrix}\right)\,,\qquad  \\
D_5\supset C_2 &:\quad
   \left(\begin{smallmatrix} 2 & 2 & 4 & 1 & 1 \\
                             \c & 1 & \c & 1 & 1
          \end{smallmatrix}\right)\,.\qquad  
\end{alignat*}
We give here some examples of branching rules for the $D_5\supset A_3\times 2A_1$ and $D_5\supset C_2$ cases, for orbits of size 10, 40 and 16 respectively, together with their corresponding indices $\gamma$.
\begin{alignat*}{1}
 D_5\supset &~ A_3\times 2A_1: \\
 & (a,0,0,0,0)\supset (0,a,0)(0)(0)+(0,0,0)(a)(a)\,, \\
 & (0,b,0,0,0)\supset (0,b,0)(b)(b)+(b,0,b)(0)(0)+(0,0,0)(2b)(0)+(0,0,0)(0)(2b)\,, \\
 & (0,0,0,0,c)\supset (0,0,c)(c)(0)+(c,0,0)(0)(c) \,, \\
&\gamma=1\,,\\
D_5\supset &~C_2: \\
 & (a,0,0,0,0)\supset (2a,0)+(0,a)+2(0,0)\,, \\
 & (0,b,0,0,0)\supset (2b,b)+(0,2b)+3(2b,0)+4(0,b)\,, \\
 & (0,0,0,0,c)\supset  (c,c)+2(c,0)\,, \\
&\gamma=5/6\,.\\
\end{alignat*}

\subsection{Rank 6}\
We give the projection matrices of the nine cases to consider. Examples of branching rules for the first six cases can be found in the corresponding subsections of the general rank section 5.7.
\begin{alignat*}{2}
D_6\supset A_5\times U_1 &: \left(
\begin{smallmatrix} 1 & 1 & \c & \c & \c & \c \\
                   \c & \c & 1 & 1 & \c & \c \\ 
                   \c & \c & \c & \c & \c & 1 \\
                   \c & \c & \c & 1 & 1 & \c \\ 
                   \c & 1 & 1 & \c & \c & \c \\
                   1 & \c & 1 & \c & 1 & \c
\end{smallmatrix}\right)\,,\qquad 
&D_6\supset D_5\times U_1 &:\left(
\begin{smallmatrix} 1 & \c & \c & \c & \c & \c \\
                   \c & 1 & \c & \c & \c & \c \\ 
                   \c & \c & 1 & \c & \c & \c \\
                   \c & \c & \c & 1 & \c & \c \\ 
                   \c & \c & \c & 1 & 1 & 1 \\
                   \c & \c & \c & \c & 1 & -1
\end{smallmatrix}\right)\,,\qquad  \\
D_6\supset B_5 &: \left(
\begin{smallmatrix} 1 & \c & \c & \c & \c & \c \\
                   \c & 1 & \c & \c & \c & \c \\ 
                   \c & \c & 1 & \c & \c & \c \\
                   \c & \c & \c & 1 & \c & \c \\ 
                   \c & \c & \c & \c & 1 & 1
\end{smallmatrix}\right)\,,\qquad 
&D_6\supset B_4\times A_1 &: \left(
\begin{smallmatrix} 
                   1 & \c & \c & \c & \c & \c \\ 
                   \c & 1 & \c & \c & \c & \c \\
                   \c & \c & 1 & \c & \c & \c \\ 
                   \c & \c & \c & 2 & 1 & 1 \\
                   \c & \c & \c & \c & 1 & 1 
\end{smallmatrix}\right)\,,\qquad \\
D_6\supset B_3\times C_2 &: \left(
\begin{smallmatrix} 
                   1 & \c & \c & \c & \c & \c \\
                   \c & 1 & 1 & \c & \c & \c \\ 
                   \c & \c & \c & 2 & 1 & 1 \\
                   \c & \c & \c & \c & 1 & 1 \\
                   \c & \c & 1 & 1 & \c & \c 
\end{smallmatrix}\right)\,,\qquad 
&D_6\supset D_4\times 2A_1 &: \left(
\begin{smallmatrix} 
                   1 & \c & \c & \c & \c & \c \\
                   \c & 1 & 1 & \c & \c & \c \\ 
                   \c & \c & \c & 1 & 1 & \c \\ 
                   \c & \c & \c & 1 & \c & 1 \\
                   \c & \c & 1 & 1 & 1 & 1 \\ 
                   \c & \c & 1 & 1 & \c & \c 
\end{smallmatrix}\right)\,,\qquad \\
D_6\supset 2A_3 &:\left(
\begin{smallmatrix} \c & \c & \c & 1 & 1 & \c \\
                   \c & 1 & 1 & \c & \c & \c \\ 
                   \c & \c & \c & 1 & \c & 1 \\
                   \c & \c & 1 & 1 & \c & \c \\ 
                   1 & 1 & \c & \c & \c & \c \\ 
                   \c & \c & 1 & 1 & 1 & 1
\end{smallmatrix}\right)\,,\qquad 
&D_6\supset 3A_1 &:\left(
\begin{smallmatrix} 2 & 4 & 6 & 6 & 4 & 4 \\
                   1 & 2 & 1 & 2 & \c & 1 \\ 
                   1 & \c & 1 & 2 & 1 & \c
\end{smallmatrix}\right)\,,\qquad \\
D_6\supset C_3\times A_1 &:\left(
\begin{smallmatrix} 
                   1 & \c & 1 & 1 & \c & \c \\ 
                   \c & 1 & 1 & \c & \c & 1 \\
                   \c & \c & \c & 1 & 1 & \c \\
                   1 & 2 & 1 & 2 & \c & 1 
\end{smallmatrix}\right)\,.\qquad 
\end{alignat*}
We give here some examples of branching rules for the last three cases, for orbits of size 12, 60 and 32 respectively, together with their corresponding indices $\gamma$.
\begin{alignat*}{1}
 D_6\supset &~ 2A_3: \\
 & (a,0,0,0,0,0)\supset (0,0,0)(0,a,0)+(0,a,0)(0,0,0)\,, \\
 & (0,b,0,0,0,0)\supset (0,b,0)(0,b,0)+(0,0,0)(b,0,b)+(b,0,b)(0,0,0)\,, \\
 & (0,0,0,0,0,c)\supset (0,0,c)(0,0,c)+(c,0,0)(c,0,0) \,, \\
&\gamma=1\,,
\end{alignat*}
\begin{alignat*}{1}
D_6\supset &~ 3A_1: \\
& (a,0,0,0,0,0)\supset (2a)(a)(a)+(0)(a)(a)\,, \\
 & (0,b,0,0,0,0)\supset (4b)(2b)(0)+(4b)(0)(2b)+2(4b)(0)(0)+(2b)(2b)(2b)\\
 & \qquad\qquad\qquad\quad +2(2b)(2b)(0)+2(2b)(0)(2b)+(0)(2b)(2b)+4(2b)(0)(0)\\
 & \qquad\qquad\qquad\quad +3(0)(2b)(0)+3(0)(0)(2b)\,, \\
 & (0,0,0,0,0,c)\supset  (4c)(c)(0)+(2c)(c)(2c)+2(2c)(c)(0)+(0)(3c)(0)\\
 & \qquad\qquad\qquad\quad +(0)(c)(2c)+3(0)(c)(0)\,, \\
&\gamma=3/7\,,\\
D_6\supset &~ C_3\times A_1: \\
& (a,0,0,0,0,0)\supset (a,0,0)(a)\,, \\
 & (0,b,0,0,0,0)\supset (0,b,0)(2b)+(2b,0,0)(0)+2(0,b,0)(0)+3(0,0,0)(2b)\,, \\
 & (0,0,0,0,0,c)\supset  (0,c,0)(c)+(0,0,0)(3c)+3(0,0,0)(c)\,, \\
&\gamma=1\,.
\end{alignat*}

\subsection{Rank 7}\
We give the projection matrices of the eleven cases to consider. Examples of branching rules for the first eight cases can be found in the corresponding subsections of the general rank section 5.7.
\begin{alignat*}{2}
D_7\supset A_6\times U_1 &: \left(
\begin{smallmatrix} 1 & 1 & \c & \c & \c & \c & \c \\
                   \c & \c & 1 & 1 & \c & \c & \c \\
                   \c & \c & \c & \c & 1 & 1 & \c \\ 
                   \c & \c & \c & \c & \c & \c & 1 \\ 
                   \c & \c & \c & 1 & 1 & \c & \c \\ 
                   \c & 1 & 1 & \c & \c & \c & \c \\
                   2 & \c & 2 & \c & 2 & -1 & 1
\end{smallmatrix}\right)\,,\qquad 
&D_7\supset D_6\times U_1&: \left(
\begin{smallmatrix} 1 & \c & \c & \c & \c & \c & \c \\
                   \c & 1 & \c & \c & \c & \c & \c \\
                   \c & \c & 1 & \c & \c & \c & \c \\ 
                   \c & \c & \c & 1 & \c & \c & \c \\ 
                   \c & \c & \c & \c & 1 & \c & \c \\ 
                   \c & \c & \c & \c & 1 & 1 & 1 \\
                   \c & \c & \c & \c & \c & 1 & -1
\end{smallmatrix}\right)\,,\qquad \\
D_7\supset B_6 &: \left(
\begin{smallmatrix} 1 & \c & \c & \c & \c & \c & \c \\
                   \c & 1 & \c & \c & \c & \c & \c \\
                   \c & \c & 1 & \c & \c & \c & \c \\ 
                   \c & \c & \c & 1 & \c & \c & \c \\ 
                   \c & \c & \c & \c & 1 & \c & \c \\ 
                   \c & \c & \c & \c & \c & 1 & 1
\end{smallmatrix}\right)\,,\qquad 
&D_7\supset B_5\times A_1 &: \left(
\begin{smallmatrix} 
                   1 & \c & \c & \c & \c & \c & \c \\
                   \c & 1 & \c & \c & \c & \c & \c \\ 
                   \c & \c & 1 & \c & \c & \c & \c \\ 
                   \c & \c & \c & 1 & \c & \c & \c \\ 
                   \c & \c & \c & \c & 2 & 1 & 1 \\
                   \c & \c & \c & \c & \c & 1 & 1 
\end{smallmatrix}\right)\,,\qquad \\
D_7\supset B_4\times C_2 &: \left(
\begin{smallmatrix}
                   1 & \c & \c & \c & \c & \c & \c \\ 
                   \c & 1 & \c & \c & \c & \c & \c \\ 
                   \c & \c & 1 & 1 & \c & \c & \c \\ 
                   \c & \c & \c & \c & 2 & 1 & 1 \\
                    \c & \c & \c & \c & \c & 1 & 1 \\
                   \c & \c & \c & 1 & 1 & \c & \c 
\end{smallmatrix}\right)\,,\qquad 
&D_7\supset 2B_3 &: \left(
\begin{smallmatrix} 1 & 1 & \c & \c & \c & \c & \c \\
                   \c & \c & 1 & 1 & \c & \c & \c \\
                   \c & \c & \c & \c & 2 & 1 & 1 \\ 
                   \c & 1 & 1 & \c & \c & \c & \c \\ 
                   \c & \c & \c & 1 & 1 & \c & \c \\ 
                   \c & \c & \c & \c & \c & 1 & 1
\end{smallmatrix}\right)\,,\qquad \\
D_7\supset D_5\times 2A_1 &: \left(
\begin{smallmatrix}  1 & \c & \c & \c & \c & \c & \c \\ 
                   \c & 1 & \c & \c & \c & \c & \c \\ 
                   \c & \c & 1 & 1 & \c & \c & \c \\ 
                   \c & \c & \c & \c & 1 & 1 & \c \\ 
                   \c & \c & \c & \c & 1 & \c & 1 \\
                   \c & \c & \c & 1 & 1 & 1 & 1 \\
                   \c & \c & \c & 1 & 1 & \c & \c 
\end{smallmatrix}\right)\,,\qquad 
&D_7\supset D_4\times A_3 &: \left(
\begin{smallmatrix} 
                   1 & 1 & \c & \c & \c & \c & \c \\ 
                   \c & \c & 1 & 1 & \c & \c & \c \\ 
                   \c & \c & \c & \c & 1 & 1 & \c \\ 
                   \c & \c & \c & \c & 1 & \c & 1 \\
                   \c & \c & \c & 1 & 1 & \c & \c \\
                   \c & 1 & 1 & \c & \c & \c & \c \\
                   \c & \c & \c & 1 & 1 & 1 & 1 
\end{smallmatrix}\right)\,,\qquad \\
D_7\supset C_2 &: \left(
\begin{smallmatrix} \c & 2 & 2 & 6 & 4 & 3 & 3 \\
                   2 & 2 & 3 & 1 & 3 & 1 & 1
\end{smallmatrix}\right)\,,\qquad 
&D_7\supset C_3 &: \left(
\begin{smallmatrix} \c & 1 & \c & 1 & \c & 1 & 1 \\
                   1 & \c & \c & 1 & 3 & 1 & 1 \\
                   \c & 1 & 2 & 1 & \c & \c & \c
\end{smallmatrix}\right)\,,\qquad \\
D_7\supset G_2 &:\left(
\begin{smallmatrix} 1 & \c & \c & 1 & \c & 1 & 1 \\
                   \c & 3 & 4 & 3 & 5 & 1 & 1
\end{smallmatrix}\right)\,.\qquad 
\end{alignat*}
We give here some examples of branching rules for the last three cases, for orbits of size 14, 84 and 64 respectively, together with their corresponding indices $\gamma$.
\begin{alignat*}{1}
 D_7\supset &~ C_2 :\\
 & (a,0,0,0,0,0,0)\supset (0,2a)+(2a,0)+(0,a)+2(0,0)\,, \\
 & (0,b,0,0,0,0,0)\supset (2b,2b)+(0,3b)+2(2b,b)+(4b,0)+3(0,2b)+5(2b,0) \\
 & \qquad\qquad\qquad\qquad  +5(0,b)\,, \\
 & (0,0,0,0,0,0,c)\supset (3c,c)+(c,2c)+2(3c,0)+3(c,c)+4(c,0) \,, \\
&\gamma=1/2\,,
\end{alignat*}
\begin{alignat*}{1}
D_7\supset &~C_3 : \\
 & (a,0,0,0,0,0,0)\supset (0,a,0)+2(0,0,0)\,, \\
 & (0,b,0,0,0,0,0)\supset (b,0,b)+2(2b,0,0)+4(0,b,0)\,, \\
 & (0,0,0,0,0,0,c)\supset  (c,c,0)+2(0,0,c)+4(c,0,0)\,, \\
&\gamma=7/6\,,\\
 D_7\supset &~G_2 : \\
 & (a,0,0,0,0,0,0)\supset (a,0)+(0,a)+2(0,0)\,, \\
 & (0,b,0,0,0,0,0)\supset (b,b)+(0,3b)+2(0,2b)+4(b,0)+5(0,b)\,, \\
 & (0,0,0,0,0,0,c)\supset (c,c)+2(0,2c)+2(c,0)+4(0,c)+4(0,0) \,, \\
&\gamma=7/8\,.\\
\end{alignat*}

\subsection{Rank 8}\
We give the projection matrices of the twelve cases to consider. Examples of branching rules for the first nine cases can be found in the corresponding subsections of the general rank section 5.7.
\begin{alignat*}{2}
D_8\supset A_7\times U_1 &: \left(
\begin{smallmatrix} 1 & 1 & \c & \c & \c & \c & \c & \c \\       
                   \c & \c & 1 & 1 & \c & \c & \c & \c \\
                   \c & \c & \c & \c & 1 & 1 & \c & \c \\
                   \c & \c & \c & \c & \c & \c & \c & 1 \\
                   \c & \c & \c & \c & \c & 1 & 1 & \c \\ 
                   \c & \c & \c & 1 & 1 & \c & \c & \c \\
                   \c & 1 & 1 & \c & \c & \c & \c & \c \\
                   1 & \c & 1 & \c & 1 & \c & 1 & \c
\end{smallmatrix}\right)\,,\qquad 
&D_8\supset D_7\times U_1 &: \left(
\begin{smallmatrix} 1 & \c & \c & \c & \c & \c & \c & \c \\       
                   \c & 1 & \c & \c & \c & \c & \c & \c \\
                   \c & \c & 1 & \c & \c & \c & \c & \c \\
                   \c & \c & \c & 1 & \c & \c & \c & \c \\
                   \c & \c & \c & \c & 1 & \c & \c & \c \\ 
                   \c & \c & \c & \c & \c & 1 & \c & \c \\
                   \c & \c & \c & \c & \c & 1 & 1 & 1 \\
                   \c & \c & \c & \c & \c & \c & 1 & -1
\end{smallmatrix}\right)\,,\qquad \\
D_8\supset B_7 &: \left(
\begin{smallmatrix} 1 & \c & \c & \c & \c & \c & \c & \c \\       
                   \c & 1 & \c & \c & \c & \c & \c & \c \\
                   \c & \c & 1 & \c & \c & \c & \c & \c \\
                   \c & \c & \c & 1 & \c & \c & \c & \c \\
                   \c & \c & \c & \c & 1 & \c & \c & \c \\ 
                   \c & \c & \c & \c & \c & 1 & \c & \c \\
                   \c & \c & \c & \c & \c & \c & 1 & 1
\end{smallmatrix}\right)\,,\qquad 
&D_8\supset B_6\times A_1 &: \left(
\begin{smallmatrix}       
                   1 & \c & \c & \c & \c & \c & \c & \c \\
                   \c & 1 & \c & \c & \c & \c & \c & \c \\
                   \c & \c & 1 & \c & \c & \c & \c & \c \\
                   \c & \c & \c & 1 & \c & \c & \c & \c \\ 
                   \c & \c & \c & \c & 1 & \c & \c & \c \\
                   \c & \c & \c & \c & \c & 2 & 1 & 1 \\
                   \c & \c & \c & \c & \c & \c & 1 & 1 
\end{smallmatrix}\right)\,,\qquad \\
D_8\supset B_5\times C_2 &: \left(
\begin{smallmatrix} 
                   1 & \c & \c & \c & \c & \c & \c & \c \\
                   \c & 1 & \c & \c & \c & \c & \c & \c \\
                   \c & \c & 1 & \c & \c & \c & \c & \c \\ 
                   \c & \c & \c & 1 & 1 & \c & \c & \c \\
                   \c & \c & \c & \c & \c & 2 & 1 & 1 \\
                   \c & \c & \c & \c & \c & \c & 1 & 1 \\       
                   \c & \c & \c & \c & 1 & 1 & \c & \c 
\end{smallmatrix}\right)\,,\qquad 
&D_8\supset B_4\times B_3 &: \left(
\begin{smallmatrix} 
                   1 & \c & \c & \c & \c & \c & \c & \c \\
                   \c & 1 & 1 & \c & \c & \c & \c & \c \\ 
                   \c & \c & \c & 1 & 1 & \c & \c & \c \\
                   \c & \c & \c & \c & \c & 2 & 1 & 1 \\
                   \c & \c & 1 & 1 & \c & \c & \c & \c \\       
                   \c & \c & \c & \c & 1 & 1 & \c & \c \\
                   \c & \c & \c & \c & \c & \c & 1 & 1 
\end{smallmatrix}\right)\,,\qquad \\
D_8\supset D_6\times 2A_1 &: \left(
\begin{smallmatrix} 
                   1 & \c & \c & \c & \c & \c & \c & \c \\
                   \c & 1 & \c & \c & \c & \c & \c & \c \\
                   \c & \c & 1 & \c & \c & \c & \c & \c \\ 
                   \c & \c & \c & 1 & 1 & \c & \c & \c \\
                   \c & \c & \c & \c & \c & 1 & 1 & \c \\
                   \c & \c & \c & \c & \c & 1 & \c & 1 \\
                   \c & \c & \c & \c & 1 & 1 & 1 & 1 \\       
                   \c & \c & \c & \c & 1 & 1 & \c & \c \\
\end{smallmatrix}\right)\,,\qquad 
&D_8\supset D_5\times A_3 &: \left(
\begin{smallmatrix} 
                   1 & \c & \c & \c & \c & \c & \c & \c \\
                   \c & 1 & 1 & \c & \c & \c & \c & \c \\ 
                   \c & \c & \c & 1 & 1 & \c & \c & \c \\
                   \c & \c & \c & \c & \c & 1 & 1 & \c \\
                   \c & \c & \c & \c & \c & 1 & \c & 1 \\
                   \c & \c & \c & \c & 1 & 1 & \c & \c \\       
                   \c & \c & 1 & 1 & \c & \c & \c & \c \\
                   \c & \c & \c & \c & 1 & 1 & 1 & 1 
\end{smallmatrix}\right)\,,\qquad \\
D_8\supset 2D_4 &: \left(
\begin{smallmatrix} 1 & 1 & \c & \c & \c & \c & \c & \c \\       
                   \c & \c & 1 & 1 & \c & \c & \c & \c \\
                   \c & \c & \c & \c & 1 & 1 & \c & \c \\
                   \c & \c & \c & \c & 1 & 1 & 1 & 1 \\
                   \c & 1 & 1 & \c & \c & \c & \c & \c \\ 
                   \c & \c & \c & 1 & 1 & \c & \c & \c \\
                   \c & \c & \c & \c & \c & 1 & 1 & \c \\
                   \c & \c & \c & \c & \c & 1 & \c & 1
\end{smallmatrix}\right)\,,\qquad 
&D_8\supset B_4 &:\left(
\begin{smallmatrix} \c & \c & \c & 1 & 1 & \c & 1 & \c \\       
                   \c & \c & 1 & \c & 1 & 1 & \c & \c \\
                   \c & 1 & \c & \c & \c & \c & \c & 1 \\
                   1 & \c & 1 & 2 & 1 & 2 & 1 & \c
\end{smallmatrix}\right)\,,\qquad \\
D_8\supset 2C_2 &: \left(
\begin{smallmatrix} 1 & 2 & 1 & 2 & 1 & 2 & 2 & 1 \\       
                   \c & \c & 1 & 1 & 2 & 1 & \c & 1 \\
                   1 & \c & 1 & 2 & 1 & 2 & \c & 1 \\
                   \c & 1 & 1 & \c & 1 & 1 & 1 & \c
\end{smallmatrix}\right)\,,\qquad 
&D_8\supset C_4\times A_1 &: \left(
\begin{smallmatrix}       
                   1 & \c & 1 & 1 & \c & \c & \c & \c \\
                   \c & 1 & 1 & \c & 1 & 1 & \c & \c \\
                   \c & \c & \c & 1 & 1 & \c & \c & 1 \\
                   \c & \c & \c & \c & \c & 1 & 1 & \c \\
                   1 & 2 & 1 & 2 & 1 & 2 & \c & 1 
\end{smallmatrix}\right)\,.\qquad 
\end{alignat*}
We give here some examples of branching rules for the last three cases, for orbits of size 16, 112 and 128 respectively, together with their corresponding indices $\gamma$.
\begin{alignat*}{1}
 D_8\supset &~B_4 :\\
 & (a,0,0,0,0,0,0,0)\supset (0,0,0,a)\,, \\
 & (0,b,0,0,0,0,0,0)\supset (0,0,b,0)+2(0,b,0,0)+4(b,0,0,0)\,, \\
 & (0,0,0,0,0,0,0,c)\supset  (0,0,c,0)+2(0,c,0,0)+(2c,0,0,0)+4(c,0,0,0) \\
 &  \qquad\qquad\qquad\qquad\quad +8(0,0,0,0)\,, \\
&\gamma=1\,,\\
D_8\supset &~2C_2 : \\
 & (a,0,0,0,0,0,0,0)\supset (a,0)(a,0)\,, \\
 & (0,b,0,0,0,0,0,0)\supset (2b,0)(0,b)+(0,b)(2b,0)+2(0,b)(0,b)+2(2b,0)(0,0)\\
 &  \qquad\qquad\qquad\qquad\quad +2(0,0)(2b,0)+4(0,b)(0,0)+4(0,0)(0,b)\,, \\
 & (0,0,0,0,0,0,0,c)\supset  (c,c)(c,0)+(c,0)(c,c)+4(c,0)(c,0)\,, \\
&\gamma=1\,,\\
D_8\supset &~C_4\times A_1: \\
 & (a,0,0,0,0,0,0,0)\supset (a,0,0,0)(a)\,, \\
 & (0,b,0,0,0,0,0,0)\supset (0,b,0,0)(2b)+(2b,0,0,0)(0)+2(0,b,0,0)(0)\\
&  \qquad\qquad\qquad\qquad\quad  +4(0,0,0,0)(2b)\,, \\
 & (0,0,0,0,0,0,0,c)\supset  (0,0,c,0)(c)+(c,0,0,0)(3c)+3(c,0,0,0)(c)\,, \\
&\gamma=1\,.\\
\end{alignat*}

\subsection{The general rank cases}\
In this section we consider infinite series of cases where the ranks of the Lie algebras take all the consecutive values starting from a lowest one. For each case, we give the corresponding projection matrix and some examples of branching rules. When the maximal reductive subalgebra of $D_n$ is semisimple, we provide also its index $\gamma$ in the Lie algebra $D_n$.

\subsubsection{$D_{2n}\supset A_{2n-1}\times U_{1}, \quad n\geq2$}

\begin{gather*}
\left(
\begin{array}{ccccccccccccc}
\text{{\tiny1}}&\text{{\tiny1}}&\c&\c&\c&\c&\dots&\c&\c&\c&\c&\c&\c\\[-1.5 ex]
\c&\c&\text{{\tiny1}}&\text{{\tiny1}}&\c&\c&{}&\c&\c&\c&\c&\c&\c\\[-1.5 ex]
{}&{}&{}&{}&{}&{}&\vdots&{}&{}&{}&{}&{}&{}\\[-1.5 ex]
\c&\c&\c&\c&\c&\c&{}&\c&\c&\text{{\tiny1}}&\text{{\tiny1}}&\c&\c\\[-1.5 ex]
\c&\c&\c&\c&\c&\c&\dots&\c&\c&\c&\c&\c&\text{{\tiny1}}\\[-1.5 ex]
\c&\c&\c&\c&\c&\c&{}&\c&\c&\c&\text{{\tiny1}}&\text{{\tiny1}}&\c\\[-1.5 ex]
\c&\c&\c&\c&\c&\c&{}&\c&\text{{\tiny1}}&\text{{\tiny1}}&\c&\c&\c\\[-1.5 ex]
{}&{}&{}&{}&{}&{}&\vdots&{}&{}&{}&{}&{}&{}\\[-1.5 ex]
\c&\text{{\tiny1}}&\text{{\tiny1}}&\c&\c&\c&\dots&\c&\c&\c&\c&\c&\c\\[-1.5 ex]
\text{{\tiny1}}&\c&\text{{\tiny1}}&\c&\text{{\tiny1}}&\c&{}&\text{{\tiny1}}&\c&\text{{\tiny1}}&\c&\text{{\tiny1}}&\c
\end{array}
\right)
\end{gather*}

\begin{align*}
(a,0,0,\dots,0)  &\supset (a,0,\dots,0)(a)+(0,\dots,0,a)(-a) \\
(0,b,0,\dots,0)  &\supset (b,0,\dots,0,b)(0)+(0,b,0,\dots,0)(2b)+(0,\dots,0,b,0)(-2b) \\
(0,0,\dots,0,c)&\supset  \left\{
\begin{array}{l l}   
(\underbrace {0,\dots,0}_{n{-}1},c,\underbrace {0,\dots,0}_{n{-}1})(0)
+(\underbrace {0,\dots,0}_{n{+}1},c,\underbrace {0,\dots,0}_{n{-}3})(2c) &\\
+(\underbrace {0,\dots,0}_{n{-}3},c,\underbrace {0,\dots,0}_{n{+}1})({-}2c)
+(\underbrace {0,\dots,0}_{n{+}3},c,\underbrace {0,\dots,0}_{n{-}5})(4c) &\\
+(\underbrace {0,\dots,0}_{n{-}5},c,\underbrace {0,\dots,0}_{n{+}3})({-}4c)+... &\\
+(0,\dots,0,c,0)((n{-}2)c)+(0,c,0,\dots,0)({-}(n{-}2)c)&\\
+(0,\dots,0)(nc)+(0,\dots,0)({-}nc) & \text{$n$ even}\\
\\
\\
(\underbrace {0,\dots,0}_{n{-}1},c,\underbrace {0,\dots,0}_{n{-}1})(0)
+(\underbrace {0,\dots,0}_{n{+}1},c,\underbrace {0,\dots,0}_{n{-}3})(2c) &\\
+(\underbrace {0,\dots,0}_{n{-}3},c,\underbrace {0,\dots,0}_{n{+}1})({-}2c)
+(\underbrace {0,\dots,0}_{n{+}3},c,\underbrace {0,\dots,0}_{n{-}5})(4c) &\\
+(\underbrace {0,\dots,0}_{n{-}5},c,\underbrace {0,\dots,0}_{n{+}3})({-}4c)+... 
+(0,\dots,0,c)((n{-}1)c)&\\
+(c,0,\dots,0)({-}(n{-}1)c) & \text{$n$ odd}\\
\end{array}\right.
\end{align*}

\subsubsection{$D_{2n+1}\supset A_{2n}\times U_{1}, \quad n\geq2$}

\begin{gather*}
\left(
\begin{array}{cccccccccccccc}
\text{{\tiny1}}&\text{{\tiny1}}&\c&\c&\c&\c&\dots&\c&\c&\c&\c&\c&\c&\c\\[-1.5 ex]
\c&\c&\text{{\tiny1}}&\text{{\tiny1}}&\c&\c&{}&\c&\c&\c&\c&\c&\c&\c\\[-1.5 ex]
{}&{}&{}&{}&{}&{}&\vdots&{}&{}&{}&{}&{}&{}&{}\\[-1.5 ex]
\c&\c&\c&\c&\c&\c&{}&\c&\c&\c&\c&\text{{\tiny1}}&\text{{\tiny1}}&\c\\[-1.5 ex]
\c&\c&\c&\c&\c&\c&\dots&\c&\c&\c&\c&\c&\c&\text{{\tiny1}}\\[-1.5 ex]
\c&\c&\c&\c&\c&\c&{}&\c&\c&\c&\text{{\tiny1}}&\text{{\tiny1}}&\c&\c\\[-1.5 ex]
\c&\c&\c&\c&\c&\c&{}&\c&\text{{\tiny1}}&\text{{\tiny1}}&\c&\c&\c&\c\\[-1.5 ex]
{}&{}&{}&{}&{}&{}&\vdots&{}&{}&{}&{}&{}&{}&{}\\[-1.5 ex]
\c&\text{{\tiny1}}&\text{{\tiny1}}&\c&\c&\c&\dots&\c&\c&\c&\c&\c&\c&\c\\[-1.5 ex]
\text{{\tiny2}}&\c&\text{{\tiny2}}&\c&\text{{\tiny2}}&\c&{}&\text{{\tiny2}}&\c&\text{{\tiny2}}&\c&\text{{\tiny2}}&\text{{\tiny-1}}&\text{{\tiny1}}
\end{array}
\right)
\end{gather*}

\begin{align*}
(a,0,0,\dots,0)  &\supset (a,0,\dots,0)(2a)+(0,\dots,0,a)(-2a) \\
(0,b,0,\dots,0)  &\supset (b,0,\dots,0,b)(0)+(0,b,0,\dots,0)(4b)+(0,\dots,0,b,0)(-4b) \\
(0,0,\dots,0,c)  &\supset \left\{
\begin{array}{l l}   
(\underbrace {0,\dots,0}_{n},c,\underbrace {0,\dots,0}_{n-1})(c)
+(\underbrace {0,\dots,0}_{n-2},c,\underbrace {0,\dots,0}_{n+1})(-3c) &\\
+(\underbrace {0,\dots,0}_{n+2},c,\underbrace {0,\dots,0}_{n-3})(5c)
+(\underbrace {0,\dots,0}_{n-4},c,\underbrace {0,\dots,0}_{n+3})(-7c) &\\
+...
+(c,0,\dots,0)({-}(2n{-}1)c)
+(0,\dots,0)((2n{+}1)c) & \text{$n$ even}\\
\\
\\
(\underbrace {0,\dots,0}_{n},c,\underbrace {0,\dots,0}_{n-1})(c)
+(\underbrace {0,\dots,0}_{n-2},c,\underbrace {0,\dots,0}_{n+1})(-3c) &\\
+(\underbrace {0,\dots,0}_{n+2},c,\underbrace {0,\dots,0}_{n-3})(5c)
+(\underbrace {0,\dots,0}_{n-4},c,\underbrace {0,\dots,0}_{n+3})(-7c)& \\
+...
+(0,\dots,0,c)((2n{-}1)c)
+(0,\dots,0)({-}(2n{+}1)c) & \text{$n$ odd}
\end{array}\right.
\end{align*}

\subsubsection{$D_{n}\supset D_{n-1}\times U_{1}, \quad n\geq5$}

\begin{gather*}
\left(
\begin{array}{cccc}
{I_{n-3}}&\multicolumn{3}{|c}{\bf 0}\\[1ex]
\hline
\multicolumn{1}{c|}{ }&\text{{\tiny1}}&\c&\c\\[-1.5 ex]
\multicolumn{1}{c|}{\bf 0}&\text{{\tiny1}}&\text{{\tiny1}}&\text{{\tiny1}}\\[-1.5 ex]
\multicolumn{1}{c|}{}&\c&\text{{\tiny1}}&\text{{\tiny-1}}
\end{array}
\right)
\end{gather*}

\begin{align*}
(a,0,0,\dots,0)  &\supset (a,0,\dots,0)(0)+(0,\dots,0)(2a)+(0,\dots,0)(-2a) \\
(0,b,0,\dots,0)  &\supset (0,b,0,\dots,0)(0)+(b,0,\dots,0)(2b)+(b,0,\dots,0)(-2b) \\
(0,0,\dots,0,c)&\supset (0,\dots,0,c,0)(c)+(0,\dots,0,c)(-c)
\end{align*}

\subsubsection{$D_{n}\supset B_{n-1}, \quad n\geq4$}

\begin{gather*}
\left(
\begin{array}{ccc}
{I_{n-2}}&\multicolumn{2}{|c}{\bf 0}\\[1ex]
\hline
\multicolumn{1}{c|}{\bf 0}&\text{{\tiny1}}&\text{{\tiny1}}
\end{array}
\right)
\end{gather*}

\begin{alignat*}{2}
&(a,0,0,\dots,0)  &&\supset (a,0,\dots,0)+2(0,\dots,0) \\
&(0,b,0,\dots,0)  &&\supset (0,b,0,\dots,0)+2(b,0,\dots,0) \\
&(0,0,\dots,0,c) &&\supset (0,\dots,0,c) \\
&\gamma=n/(n-1) 
\end{alignat*}

\subsubsection{$D_{n}\supset B_{n-2}\times A_{1}, \quad n\geq4$}

\begin{gather*}
\left(
\begin{array}{cccc}
{I_{n-3}}&\multicolumn{3}{|c}{\bf 0}\\[1ex]
\hline
\multicolumn{1}{c|}{ }&\text{{\tiny2}}&\text{{\tiny1}}&\text{{\tiny1}}\\[-1.5 ex]
\multicolumn{1}{c|}{\bf 0}&\c&\text{{\tiny1}}&\text{{\tiny1}}
\end{array}
\right)
\end{gather*}

\begin{alignat*}{2}
&(a,0,0,\dots,0)  &&\supset (a,0,\dots,0)(0)+(0,\dots,0)(2a)+2(0,\dots,0)(0) \\
&(0,b,0,\dots,0)  &&\supset (0,b,0,\dots,0)(0)+(b,0,\dots,0)(2b)+2(b,0,\dots,0)(0)\\
& &&\quad+2(0,\dots,0)(2b) \\
&(0,0,\dots,0,c) &&\supset (0,\dots,0,c)(c) \\
&\gamma=1
\end{alignat*}

\subsubsection{$D_{n}\supset B_{n-k-1}\times B_{k}, \quad n-k-1\geq k \geq2, \quad n\geq5$}

\begin{gather*}
\left(
\begin{array}{ccccccccccccc}
{I_{n-2k-1}}&\multicolumn{12}{|c}{\bf 0}\\[1ex]
\hline
\multicolumn{1}{c|}{ }&\text{{\tiny1}}&\text{{\tiny1}}&\c&\c&\c&\c&\dots&\c&\c&\c&\c&\c\\[-1.5 ex]
\multicolumn{1}{c|}{ }&\c&\c&\text{{\tiny1}}&\text{{\tiny1}}&\c&\c&{}&\c&\c&\c&\c&\c\\[-1.5 ex]
\multicolumn{1}{c|}{ }&{}&{}&{}&{}&{}&{}&\vdots&{}&{}&{}&{}&{}\\[-1.5 ex]
\multicolumn{1}{c|}{ }&\c&\c&\c&\c&\c&\c&{}&\text{{\tiny1}}&\text{{\tiny1}}&\c&\c&\c\\[-1.5 ex]
\multicolumn{1}{c|}{ }&\c&\c&\c&\c&\c&\c&\dots&\c&\c&\text{{\tiny2}}&\text{{\tiny1}}&\text{{\tiny1}}\\[-1.5 ex]
\multicolumn{1}{c|}{\bf 0}&\c&\text{{\tiny1}}&\text{{\tiny1}}&\c&\c&\c&{}&\c&\c&\c&\c&\c\\[-1.5 ex]
\multicolumn{1}{c|}{ }&\c&\c&\c&\text{{\tiny1}}&\text{{\tiny1}}&\c&{}&\c&\c&\c&\c&\c\\[-1.5 ex]
\multicolumn{1}{c|}{ }&{}&{}&{}&{}&{}&{}&\vdots&{}&{}&{}&{}&{}\\[-1.5 ex]
\multicolumn{1}{c|}{ }&\c&\c&\c&\c&\c&\c&\dots&\c&\text{{\tiny1}}&\text{{\tiny1}}&\c&\c\\[-1.5 ex]
\multicolumn{1}{c|}{ }&\c&\c&\c&\c&\c&\c&{}&\c&\c&\c&\text{{\tiny1}}&\text{{\tiny1}}
\end{array}
\right)
\end{gather*}

\begin{alignat*}{2}
&(a,0,0,\dots,0)  &&\supset (a,0,\dots,0)(0,\dots,0)+(0,\dots,0)(a,0,\dots,0)+2(0,\dots,0)(0,\dots,0) \\
&(0,b,0,\dots,0)  &&\supset (b,0,\dots,0)(b,0,\dots,0)+(0,b,0,\dots,0)(0,\dots,0)\\
& &&\quad+(0,\dots,0)(0,b,0,\dots,0)+2(b,0,\dots,0)(0,\dots,0)\\
& &&\quad+2(0,\dots,0)(b,0,\dots,0) \\
&(0,0,\dots,0,c) &&\supset (0,\dots,0,c)(0,\dots,0,c) \\
&\gamma=n/(n-1)
\end{alignat*}

\subsubsection{$D_{n}\supset D_{n-2}\times A_{1}\times A_{1}, \quad n\geq 6$}

\begin{gather*}
\left(
\begin{array}{cccccc}
{I_{n-5}}&\multicolumn{5}{|c}{\bf 0}\\[1ex]
\hline
\multicolumn{1}{c|}{ }&\text{{\tiny1}}&\text{{\tiny1}}&\c&\c&\c\\[-1.5 ex]
\multicolumn{1}{c|}{}&\c&\c&\text{{\tiny1}}&\text{{\tiny1}}&\c\\[-1.5 ex]
\multicolumn{1}{c|}{\bf 0}&\c&\c&\text{{\tiny1}}&\c&\text{{\tiny1}}\\[-1.5 ex]
\multicolumn{1}{c|}{}&\c&\text{{\tiny1}}&\text{{\tiny1}}&\text{{\tiny1}}&\text{{\tiny1}}\\[-1.5 ex]
\multicolumn{1}{c|}{}&\c&\text{{\tiny1}}&\text{{\tiny1}}&\c&\c
\end{array}
\right)
\end{gather*}

\begin{alignat*}{2}
&(a,0,0,\dots,0)  &&\supset (a,0,\dots,0)(0)(0)+(0,\dots,0)(a)(a) \\
&(0,b,0,\dots,0)  &&\supset (0,b,0,\dots,0)(0)(0)+(b,0,\dots,0)(b)(b)+(0,\dots,0)(2b)(0)\\
& &&\quad+(0,\dots,0)(0)(2b) \\
&(0,0,\dots,0,c) &&\supset (0,\dots,0,c)(c)(0)+(0,\dots,0,c,0)(0)(c) \\
&\gamma=1
\end{alignat*}

\subsubsection{$D_{n}\supset D_{n-3}\times A_{3}, \quad n\geq7$}

\begin{gather*}
\left(
\begin{array}{cccccccc}
{I_{n-7}}&\multicolumn{7}{|c}{\bf 0}\\[1ex]
\hline
\multicolumn{1}{c|}{ }&\text{{\tiny1}}&\text{{\tiny1}}&\c&\c&\c&\c&\c\\[-1.5 ex]
\multicolumn{1}{c|}{}&\c&\c&\text{{\tiny1}}&\text{{\tiny1}}&\c&\c&\c\\[-1.5 ex]
\multicolumn{1}{c|}{}&\c&\c&\c&\c&\text{{\tiny1}}&\text{{\tiny1}}&\c\\[-1.5 ex]
\multicolumn{1}{c|}{\bf 0}&\c&\c&\c&\c&\text{{\tiny1}}&\c&\text{{\tiny1}}\\[-1.5 ex]
\multicolumn{1}{c|}{}&\c&\c&\c&\text{{\tiny1}}&\text{{\tiny1}}&\c&\c\\[-1.5 ex]
\multicolumn{1}{c|}{}&\c&\text{{\tiny1}}&\text{{\tiny1}}&\c&\c&\c&\c\\[-1.5 ex]
\multicolumn{1}{c|}{}&\c&\c&\c&\text{{\tiny1}}&\text{{\tiny1}}&\text{{\tiny1}}&\text{{\tiny1}}
\end{array}
\right)
\end{gather*}

\begin{alignat*}{2}
&(a,0,0,\dots,0)  &&\supset (a,0,\dots,0)(0,0,0)+(0,\dots,0)(0,a,0) \\
&(0,b,0,\dots,0)  &&\supset (0,b,0,\dots,0)(0,0,0)+(b,0,\dots,0)(0,b,0)+(0,\dots,0)(b,0,b) \\
&(0,0,\dots,0,c) &&\supset (0,\dots,0,c)(0,0,c)+(0,\dots,0,c,0)(c,0,0) \\
&\gamma=1
\end{alignat*}

\subsubsection{$D_{n}\supset D_{n-k}\times D_{k}, \quad n-k\geq k \geq 4$}\

\begin{gather*}
\left(
\begin{array}{ccccccccccccc}
{I_{n-2k}}&\multicolumn{12}{|c}{\bf 0}\\[1ex]
\hline
\multicolumn{1}{c|}{ }&\text{{\tiny1}}&\text{{\tiny1}}&\c&\c&\c&\c&\dots&\c&\c&\c&\c&\c\\[-1.5 ex]
\multicolumn{1}{c|}{ }&\c&\c&\text{{\tiny1}}&\text{{\tiny1}}&\c&\c&{}&\c&\c&\c&\c&\c\\[-1.5 ex]
\multicolumn{1}{c|}{ }&{}&{}&{}&{}&{}&{}&\vdots&{}&{}&{}&{}&{}\\[-1.5 ex]
\multicolumn{1}{c|}{ }&\c&\c&\c&\c&\c&\c&{}&\c&\text{{\tiny1}}&\text{{\tiny1}}&\c&\c\\[-1.5 ex]
\multicolumn{1}{c|}{ }&\c&\c&\c&\c&\c&\c&\dots&\c&\text{{\tiny1}}&\text{{\tiny1}}&\text{{\tiny1}}&\text{{\tiny1}}\\[-1.5 ex]
\multicolumn{1}{c|}{\bf 0}&\c&\text{{\tiny1}}&\text{{\tiny1}}&\c&\c&\c&{}&\c&\c&\c&\c&\c\\[-1.5 ex]
\multicolumn{1}{c|}{ }&\c&\c&\c&\text{{\tiny1}}&\text{{\tiny1}}&\c&{}&\c&\c&\c&\c&\c\\[-1.5 ex]
\multicolumn{1}{c|}{ }&{}&{}&{}&{}&{}&{}&\vdots&{}&{}&{}&{}&{}\\[-1.5 ex]
\multicolumn{1}{c|}{ }&\c&\c&\c&\c&\c&\c&\dots&\c&\c&\text{{\tiny1}}&\text{{\tiny1}}&\c\\[-1.5 ex]
\multicolumn{1}{c|}{ }&\c&\c&\c&\c&\c&\c&{}&\c&\c&\text{{\tiny1}}&\c&\text{{\tiny1}}
\end{array}
\right)
\end{gather*}

\begin{alignat*}{2}
&(a,0,0,\dots,0)  &&\supset (a,0,\dots,0)(0,\dots,0)+(0,\dots,0)(a,0,\dots,0) \\
&(0,b,0,\dots,0)  &&\supset (0,b,0,\dots,0)(0,\dots,0)+(b,0,\dots,0)(b,0,\dots,0)\\
& &&\quad+(0,\dots,0)(0,b,0,\dots,0) \\
&(0,0,\dots,0,c) &&\supset (0,\dots,0,c)(0,\dots,0,c)+(0,\dots,0,c,0)(0,\dots,0,c,0) \\
&\gamma=1
\end{alignat*}

\section{Concluding remarks}
\begin{itemize}
\item
The pairs $W(L)\supset W(L')$ in this paper involve a maximal subalgebra $L'$ in~$L$. A chain of maximal subalgebras linking $L$ and any of its reductive non-maximal subalgebras $L''$ can be found. Corresponding projection matrices combine, by common matrix multiplication, into the projection matrix for $W(L)\supset W(L'')$.
\smallskip

\item
Projection matrices of $W(L)\supset W(L')$ when the ranks of $L$ and $L'$ are the same, are square matrices with determinant different from zero. Hence they can be inverted and used in the opposite direction. The inverse matrix transforms an orbit of $W(L')$ into the linear combination of orbits of $W(L)$, where $L'\subset L$. The linear combination has integer coefficients of both signs in general. We know of no interpretation of such `branching rules' in applied literature, although they have their place in the Grothendieck rings of representations.
\smallskip

\item
Weyl group orbits retain most of their useful properties, such as decomposition of their products and branching rules, even when their points are off the weight lattice. Two applications of such orbits can be anticipated. First they could serve as models of molecules that have full Weyl group symmetry without having the rigid regularity of distances between their points/atoms. Another application is undoubtedly Fourier analysis, when Fourier integral expansions are studied rather than discrete ones.
\smallskip

\item
Curious and completely unexplored relations between pairs of maximal subalgebras, say $L'$ and $L''$, of the same Lie algebra $L$ can be found by combining the projection matrices $P(L\supset L')$ and $P(L\supset L'')$ as
\begin{gather*}
P(L'\rightarrow L'')=P(L\supset L'')P^{-1}(L\supset L').
\end{gather*}
Here $L'$ must be of the same rank as $L$ for $P(L\supset L')$ to be invertible. We write $L'\rightarrow L''$ instead of $L' \supset L''$ here because $L''$ is obviously not a subalgebra of $L'$.
\smallskip

\item
Congruence classes of representations are naturally extended to
congruence classes of $W$-orbits \cite{HLP}. Comparing the
congruence classes of orbits for $W(L)\supset W(L')$ reveals that
not all combinations of congruence classes are present. A
relative congruence class is a valid and useful concept which
deserves investigation.
\smallskip

\item
Following the experience gained from applications of finite dimensional representations of semisimple Lie algebras, one could also study, in the case of Weyl group orbits, their anomaly numbers \cite{PS81, OP85} and indices of higher than second degree \cite{PSW, OP84, McPS}. 
\smallskip

\item
Subjoining among semisimple Lie resembles inclusion because it allows one to calculate `branching rules'. Projection matrices are perfectly adequate for this task \cite{PSS}. But it is not an homomorphism, therefore it is a different relation. All maximal subjoinings have been classified \cite{MPi}.

Consider an example of subjoining. The 4-dimensional representation $(1,0,0)$ of $A_3$ does {\it not} contain the 5-dimensional representation $(0,1)$ of $C_2$. In spite of this, the projection matrix that maps the highest weight orbit of $A_3$ (and any other orbit of $A_3$) into the orbit $(0,1)$ of $C_2$ can be obtained. Indeed, that projection matrix is $\left(\begin{smallmatrix}0&2&0\\1&0&1\end{smallmatrix}\right)$.

\end{itemize}
\medskip
\medskip
\centerline{\bf Acknowledgements}
\medskip

This work was supported in part by the Natural Sciences and Engineering Research Council of Canada and by the MIND Research Institute of Santa Ana, California. M. L. is grateful for the support she receives from the Alexander Graham Bell Scholarship. 

\end{document}